\documentclass[aps,pra,10pt,amsmath,amssymb,superscriptaddress,twocolumn,notitlepage,nofootinbib,floatfix]{revtex4-2}

\usepackage[utf8]{inputenx}
\usepackage[english]{babel}
\usepackage{graphicx}
\usepackage[usenames,dvipsnames]{xcolor}
\usepackage{physics}
\usepackage{dsfont}
\usepackage{endnotes}

\definecolor{purple}{RGB}{128,0,128}
\definecolor{ultramarine}{RGB}{63, 0, 255}
\definecolor{medblue}{RGB}{0, 0, 100}
\definecolor{googleblue}{RGB}{34, 0, 204}
\definecolor{panblue}{RGB}{0,24,150}
\definecolor{carmine}{RGB}{150, 0, 24}
\definecolor{gray}{RGB}{150, 150, 150}
\definecolor{darkgreen}{RGB}{0, 80, 0}

\usepackage[]{hyperref}
\hypersetup{linkcolor=carmine,citecolor=darkgreen,urlcolor=googleblue,anchorcolor=OliveGreen}

\usepackage{microtype}
\microtypecontext{spacing=nonfrench}
\microtypesetup{
expansion={true,nocompatibility},
protrusion={true,nocompatibility},
activate={true,nocompatibility},
tracking=true,
kerning=true,
spacing={true}
}

\usepackage[style=american,autopunct=true]{csquotes} 
\usepackage{mathtools}

\usepackage{soul}

\usepackage{bbold}
\usepackage[flushleft, neveradjust]{paralist}

\usepackage[caption=false]{subfig}

\usepackage{verbatim}
\usepackage{verbatimbox}

\newcommand{\Hmin}{H_{\mathrm{min}}}

\begin{document}

\title{Experimental randomness certification in a quantum network with independent sources}

\author{Giorgio Minati}
\affiliation{Dipartimento di Fisica - Sapienza Universit\`{a} di Roma, P.le Aldo Moro 5, I-00185 Roma, Italy}
\author{Giovanni Rodari}
\affiliation{Dipartimento di Fisica - Sapienza Universit\`{a} di Roma, P.le Aldo Moro 5, I-00185 Roma, Italy}
\author{Emanuele Polino}
\affiliation{Dipartimento di Fisica - Sapienza Universit\`{a} di Roma, P.le Aldo Moro 5, I-00185 Roma, Italy}
\affiliation{Centre for Quantum Dynamics and Centre for Quantum Computation and Communication Technology Griffith University Yuggera Country Brisbane Queensland 4111 Australia}

\author{Francesco Andreoli}
\affiliation{ICFO - Institut de Ciències Fotòniques, The Barcelona Institute of Science and Technology, \\08860 Castelldefels, Spain }

\author{Davide Poderini}
\affiliation{International Institute of Physics, Federal University of Rio Grande do Norte, 59078-970, Natal, Brazil}
\affiliation{Università degli Studi di Pavia, Dipartimento di Fisica, QUIT Group, via Bassi 6, 27100 Pavia, Italy}

\author{Rafael Chaves}
\affiliation{International Institute of Physics, Federal University of Rio Grande do Norte, 59078-970, Natal, Brazil}
\affiliation{School of Science and Technology, Federal University of Rio Grande do Norte, 59078-970, Natal, Brazil}

\author{Gonzalo Carvacho}
\email[Corresponding author: ]{gonzalo.carvacho@uniroma1.it}
\affiliation{Dipartimento di Fisica - Sapienza Universit\`{a} di Roma, P.le Aldo Moro 5, I-00185 Roma, Italy}

\author{Fabio Sciarrino}
\affiliation{Dipartimento di Fisica - Sapienza Universit\`{a} di Roma, P.le Aldo Moro 5, I-00185 Roma, Italy}

\begin{abstract}

Randomness certification is a foundational and practical aspect of quantum information science, essential for securing quantum communication protocols. Traditionally, these protocols have been implemented and validated with a single entanglement source, as in the paradigmatic Bell scenario. However, advancing these protocols to support more complex configurations involving multiple entanglement sources is key to building robust architectures and realizing large-scale quantum networks. In this work, we show how to certify randomness in an entanglement-teleportation experiment, the building block of a quantum repeater displaying two independent sources of entanglement. Utilizing the scalar extension method, we address the challenge posed by the non-convexity of the correlation set, providing effective bounds on an eavesdropper's knowledge of the shared secret bits. Our theoretical model characterizes the certifiable randomness within the network and is validated through the analysis of experimental data from a photonic quantum network.

\end{abstract}

\maketitle
\section{Introduction}

Quantum non-locality has captivated scientific interest since the seminal contributions of Einstein, Podolsky, Rosen \cite{einstein1935can} and later Bell \cite{bell1964einstein}. These foundational studies have prompted extensive investigations into the limitations of local hidden variable theories, which fail to explain the predictions of quantum theory \cite{RevModPhys.86.419, doi:10.1126/science.1182103}. In parallel, advances in quantum information theory have revealed that these non-classical properties provide essential resources for practical applications such as distributed computing \cite{Beigi_2011, buhrman2010nonlocality} and cryptographic protocols \cite{RevModPhys.94.025008, Yin2020, PhysRevLett.84.4729, pirandola2020advances}.

The non-classical nature revealed by violations of Bell inequalities, serves as the foundation for secure randomness generation and certification, specifically by enabling eavesdropper-secure random bit strings through measurements on a physical system \cite{pironio2010random, nieto2014using, acin2016certified, PhysRevLett.108.100402, PhysRevA.93.040102, Woodhead_2020}. This task can be achieved in a Device-Independent (DI) framework and has been explored theoretically and experimentally \cite{Liu2018, PhysRevA.97.040102, PhysRevLett.126.050503, Shalm2021, seguinard2023experimental, Agresti2020}, predominantly within the paradigmatic Bell’s scenario, where two distant parties perform local measurements on a shared entangled resource. In this case, the secure randomness that can be generated is quantified using the concept of guessing probability \cite{colbeck2011quantum, Acin2016}, which represents the probability that an external agent, such as an eavesdropper, can correctly predict the measurement results based on the observed output statistics. Importantly, it has been shown that whenever non-classicality is manifested through the violation of a Bell inequality, a non-zero amount of randomness can be certified \cite{pironio2010random, Acin2016, PhysRevA.95.020102}. Similar studies have investigated variations of the bipartite scenario \cite{PhysRevLett.108.100402, PhysRevA.93.040102, PhysRevA.95.020102, Woodhead_2020, Nieto-Silleras_2014, Bancal_2014}, as well as other configurations, such as Bell-like \cite{Woodhead2018randomnessversus, Grasselli2023boostingdevice} or broadcasting \cite{broad2024} three-party networks and the instrumental scenario \cite{Agresti2020}. All of these scenarios share the common feature of involving a single shared source of correlations between distant parties.

Notwithstanding, identifying non-classicality in scenarios with multiple independent sources is crucial for both foundational research and practical applications in quantum technologies \cite{tavakoli2022bell, poderini2020experimental, chaves2021causal, suprano2022experimental, polino2023experimental,PRAANDREOLI, d2023machine, wang2023certification, gu2023experimental, wang2024experimental, saunders2017experimental, carvacho2022quantum}. These multi-source configurations are essential building blocks for scalable, long-range quantum communication networks. Within the network framework \cite{tavakoli2022bell}, independent sources generate a complex, non-convex set of correlations, making randomness certification particularly challenging. Consequently, existing methods for certifying randomness have shown limited effectiveness when applied to such intricate network structures \cite{Lee2018, wolfe2021quantum, sekatski2023partial}.

In this work, we address this challenge by leveraging the scalar extension technique \cite{pozas2019bounding} to establish a robust framework for randomness certification in quantum networks. To illustrate the general method, we focus on the network underlying the entanglement swapping experiment \cite{Zukowski1993}, comprised of two independent entanglement sources also known as the bilocal scenario \cite{branciard_2012}. In particular, this network structure allows for two distinct eavesdropping strategies. For both strategies, we demonstrate that source-independence enables the certification of up to $1.41$ bits of randomness between the network’s outer nodes—a figure that surpasses the $1.23$ bits certified by the maximal violation of the CHSH inequality \cite{PhysRevLett.108.100402}. Furthermore, we apply our framework to certify randomness in the experimental bilocal scenario \cite{carvacho2017experimental}, thus demonstrating the feasibility of certifying randomness against eavesdropping threats in operational quantum networks.

\section{Randomness in the Bell scenario}

\begin{figure}[t!]
\centering
\subfloat[]{\label{fig:bipartite_DAG}{\includegraphics[width=0.19\textwidth]{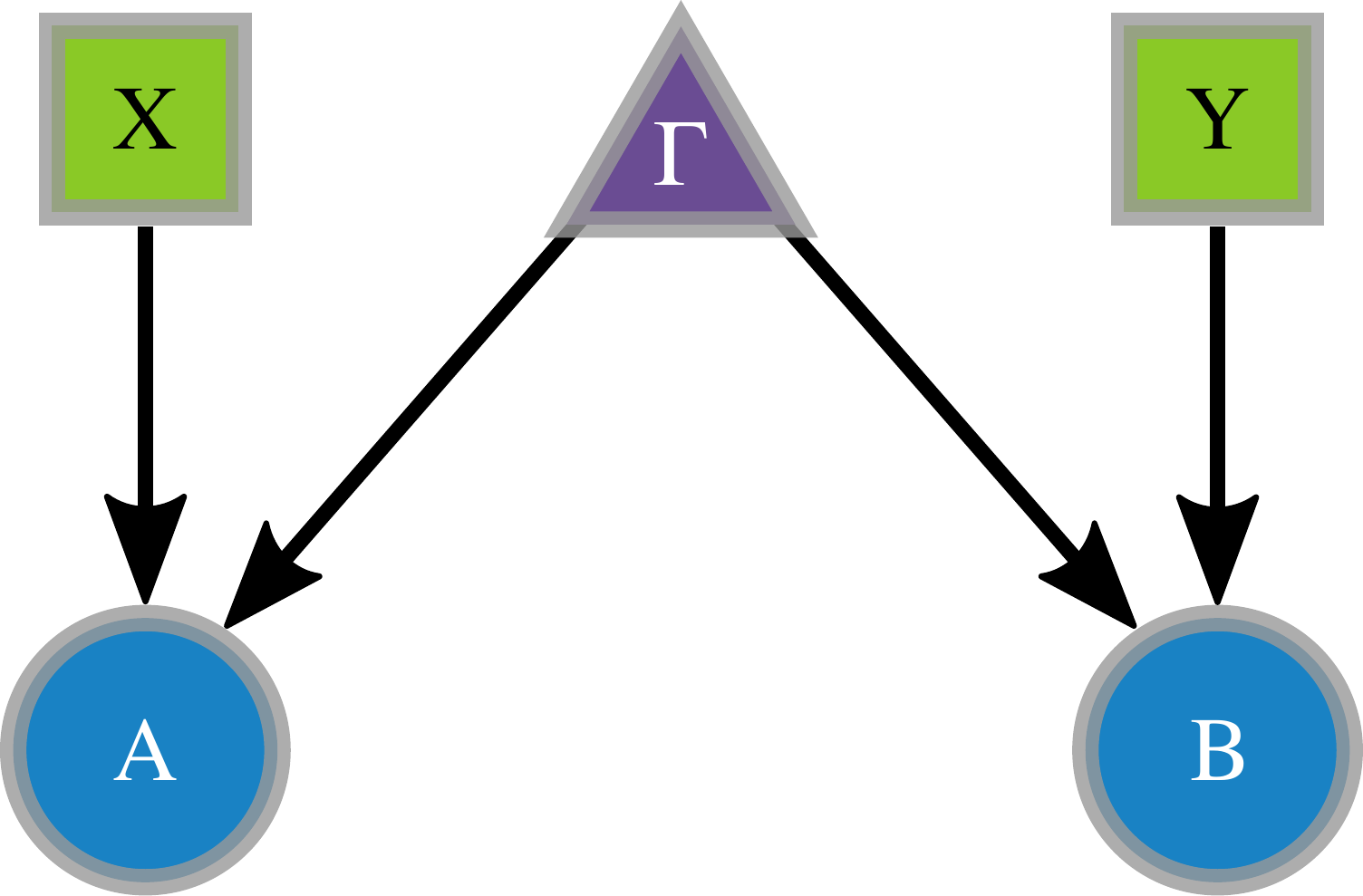}}}\hfil
\subfloat[]{\label{fig:bilocal_dag}
{\includegraphics[width=0.19\textwidth]{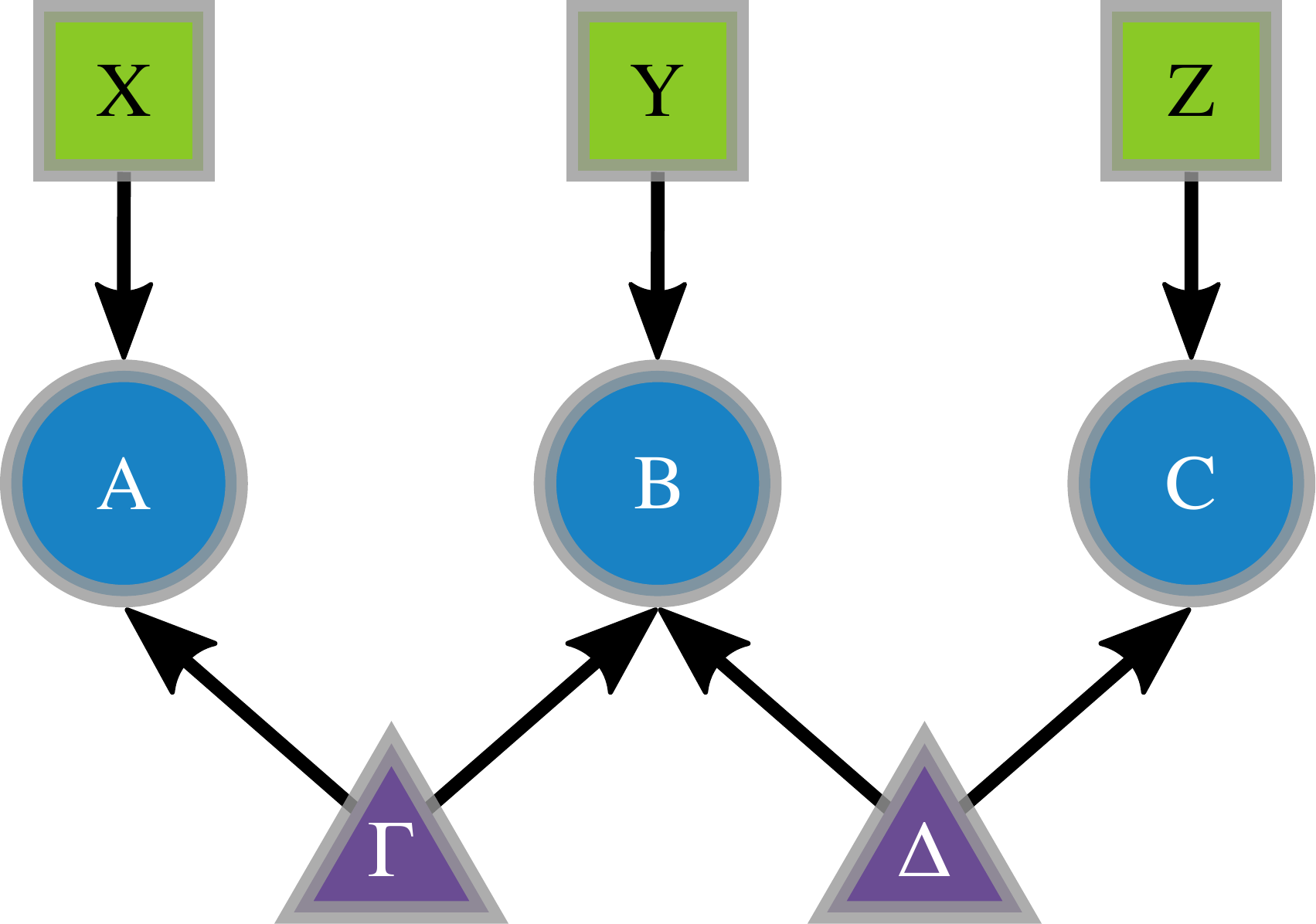}}}
\caption{\textbf{Representation of different causal structures.} Directed acyclic graphs (DAGs) represent different causal structures, the nodes in the graph represent the relevant random variables with arrows accounting for their causal relations. There are three different kinds of nodes: sources of correlations represented either by hidden variables or quantum states (purple triangles), measurement settings (green squares), and measurement outcomes (blue circles). \textbf{(a)} Bipartite model with one entangled source, \textbf{(b)} Tripartite scenario with two independent sources, accounting for the bilocal hidden variable model.}
\end{figure}

Bell's theorem \cite{brunner2014bell} is a no-go theorem proving the impossibility of reproducing the predictions of quantum theory within the classical causal model depicted in Fig. \ref{fig:bipartite_DAG}. If we consider two parties $A$ and $B$ performing local measurements on subsystems of a bipartite state $\rho_{AB}$ produced by a single common source, the output probabilities predicted by the quantum theory are
\begin{equation}
    p_Q(ab|xy) = \Tr (A_a^x \otimes B_b^y \cdot \rho_{AB})\, .
    \label{eqn:qreal}
\end{equation}
For a suitable choice of an entangled state $\rho_{AB}$ and operators $\{A_a^x, B_b^y\}$, this distribution cannot be described by the classical causal model in Fig.~\ref{fig:bipartite_DAG}, which implies its incompatibility with a hidden variable model given by
\begin{equation}
    p_L(ab|xy) = \sum_\lambda p(a|x,\lambda)p(b|y,\lambda)p(\lambda).
\end{equation}

A notable property of such non-classical distributions is the possibility to certify that the correlations established between parties $A$ and $B$ cannot be shared with a third party~\cite{acin2016certified}. Importantly, this certification relies on minimal assumptions: that an eavesdropper (E) has access to an extended quantum state $\rho_{ABE}$ and that the laboratories in which $A$ and $B$ carry out their measurements are secure. By further assuming that the eavesdropper's measurement procedure is described by Positive Operator-Valued Measure (POVM) operators $E_e$, the information accessible to the eavesdropper should arise from a joint quantum distribution
\begin{equation}
    p(abe|xy) = \Tr (A_a^x \otimes B_b^y \otimes E_e \cdot \rho_{ABE}),
    \label{eqn:ABEfull}
\end{equation}
such that $A$ and $B$ observe a specific probability distribution $p(ab|xy)$ admitting the realization of Eq.\eqref{eqn:qreal}.

To bound the amount of information that $E$ can extract over the outcomes of $A$ and $B$, one considers the {\em guessing probability}
\begin{equation}
    G(AB | E, xy) = \sum_{ab} p(ab,e=(a,b)|xy).
    \label{eqn:pguess}
\end{equation}
The amount of certifiable randomness in a certain scenario is related to the maximum of such quantity achievable with a given realization $p(ab|xy)$, a problem which can be efficiently solved through Semi-Definite Programming (SDP) techniques as the NPA (Navascués-Pironio-Acín) hierarchy \cite{Navascués_2008} under the constraint given by Eq.\eqref{eqn:ABEfull}. 
From the guessing probability, one can readily obtain the amount of certifiable randomness in bits, expressed by the so-called min-entropy \cite{PhysRevA.87.012336}, defined as:
\begin{equation}
    H_{\text{min}} = -\log_2 (G(AB|E,xy)),
\end{equation}
that can achieve values up to $1.23$ bits of randomness in the standard bipartite scenario when bounded by the CHSH inequality~\cite{pironio2010random}.
Exploring various bipartite scenarios and strategies can increase the certifiable randomness, reaching up to 2 bits per round, while decreasing its robustness to noise. This value can be obtained either by considering bipartite scenarios with additional inputs~\cite{wooltorton2023expandingbipartitebellinequalities} or, in the standard case of dichotomic inputs and outputs, using so-called ``tilted'' Bell inequalities~\cite{PhysRevLett.108.100402}. Finally, other figures of merit different from $H_{\mathrm{min}}$ have also been considered~\cite{bhavsar2023improved,wooltorton2022tight}.

\section{Randomness certification in the bilocal scenario}

Building on the concept of randomness in the standard Bell scenario, several works have addressed its variations \cite{PhysRevLett.108.100402, PhysRevA.93.040102, PhysRevA.95.020102, Woodhead_2020, Nieto-Silleras_2014, Bancal_2014} and other Bell-like scenarios of relevance \cite{Woodhead2018randomnessversus, Grasselli2023boostingdevice, Agresti2020,broad2024}. Although scenarios involving multiple independent sources of correlations are crucial for future applications, the challenge of randomness certification in these quantum networks \cite{tavakoli2022bell} remains almost unexplored \cite{Lee2018,wolfe2021quantum,sekatski2023partial}. 
In this context, the \textit{bilocal scenario} \cite{branciard_2012}, depicted in Fig.\ref{fig:bilocal_dag}, plays a prominent role since it is the underlying causal structure of entanglement swapping \cite{ent_swap_1, ent_swap_2}, an essential protocol for quantum repeaters \cite{RevModPhys.95.045006, Li2019} and long-distance communication networks \cite{Chen2021, PhysRevLett.120.030501}. It consists of two independent sources distributed among three parties: two of them receive a single subsystem coming respectively from $\rho_{AB_1}$ and $\rho_{B_2C}$, while the central node holds two independent subsystems coming from both sources. Each of the parties carries out local measurements by independently choosing among settings described by the variables $\{ X, Y, Z \}$, producing outcomes denoted as $\{ A, B, C \}$, with a probability distribution of measurements outcomes given by
\begin{equation}
\label{eq:Qbehavior_bilocal}
    p(abc|xyz) = \tr (\rho_{ABC} \cdot A_{a|x} \otimes B_{b|y} \otimes C_{c|z}),
\end{equation}
Here, $\rho_{ABC} = \rho_{AB_1} \otimes \rho_{B_2C}$ encodes the source independence and $\{A_{a|x}, B_{b|y}, C_{c|z}\}$ define the measurements described, in general, by POVM operators.

\begin{figure}[ht]
\centering
\subfloat[]{\label{fig:bilocal_dag_EF}{\includegraphics[width=0.135\textwidth]{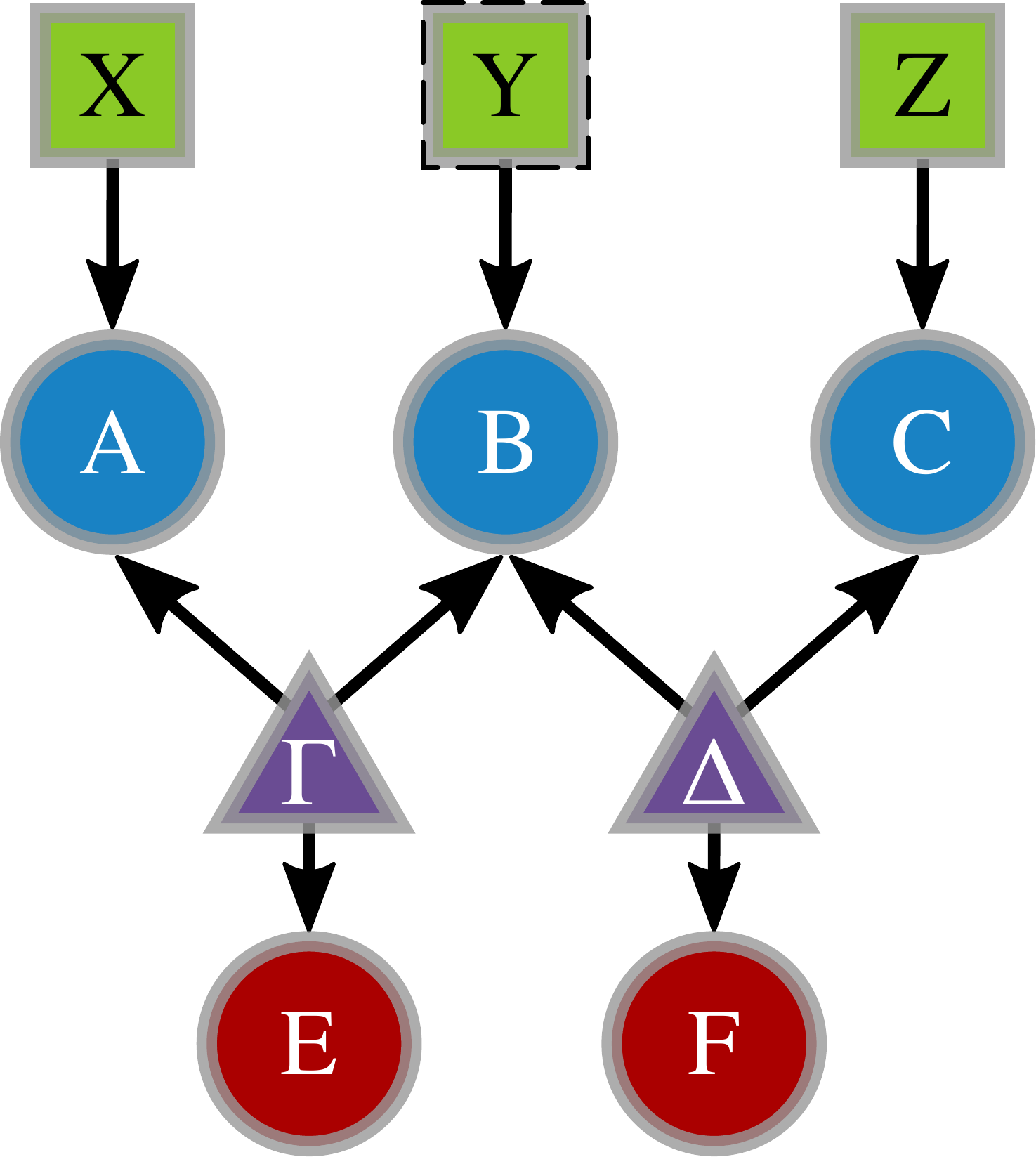}}} \hfill
\subfloat[]{\label{fig:bilocal_dag_E}{\includegraphics[width=0.135\textwidth]{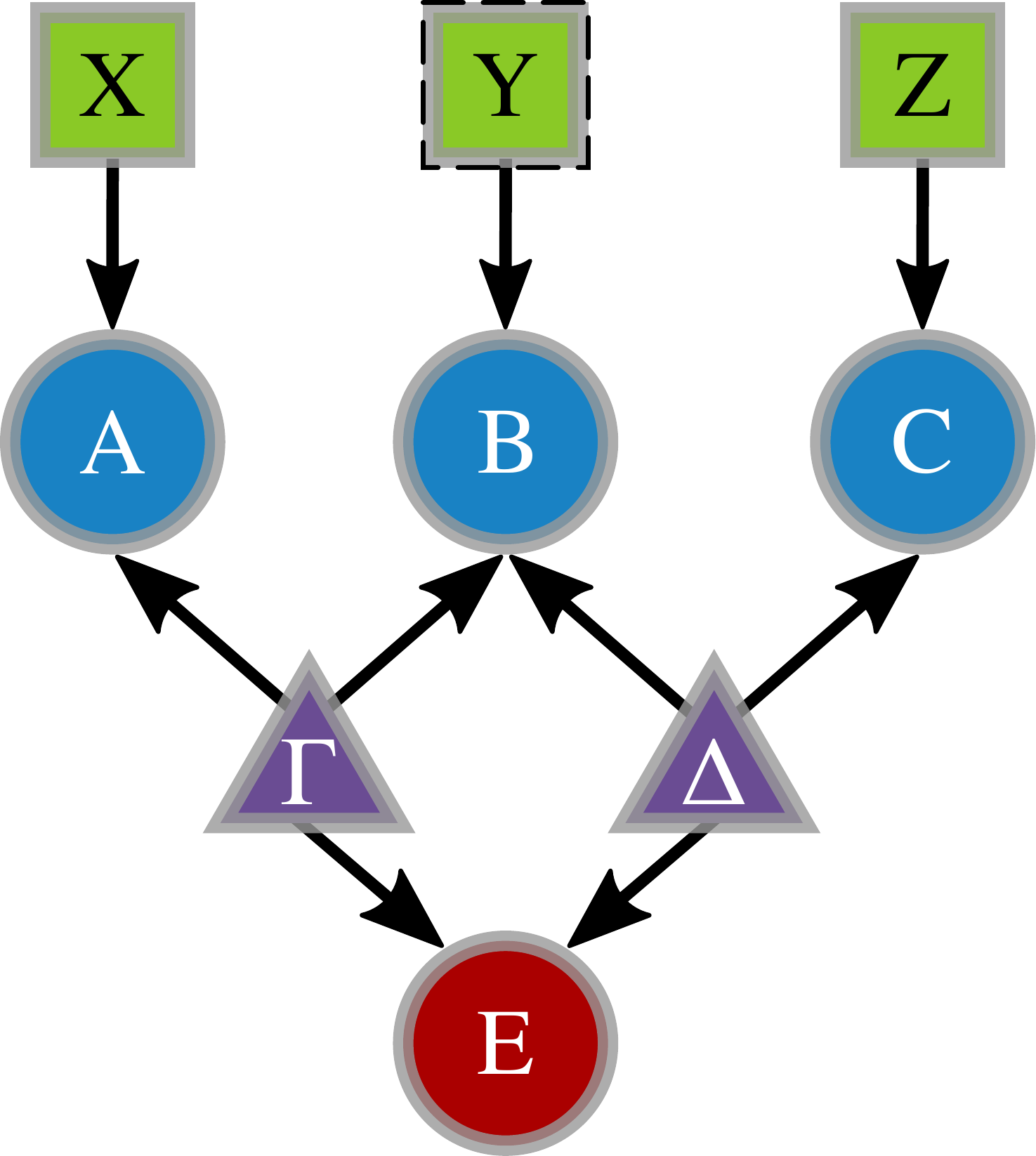}}} 
\hfill
\subfloat[]{\label{fig:bilocal_dag_ΛE}{\includegraphics[width=0.135\textwidth]{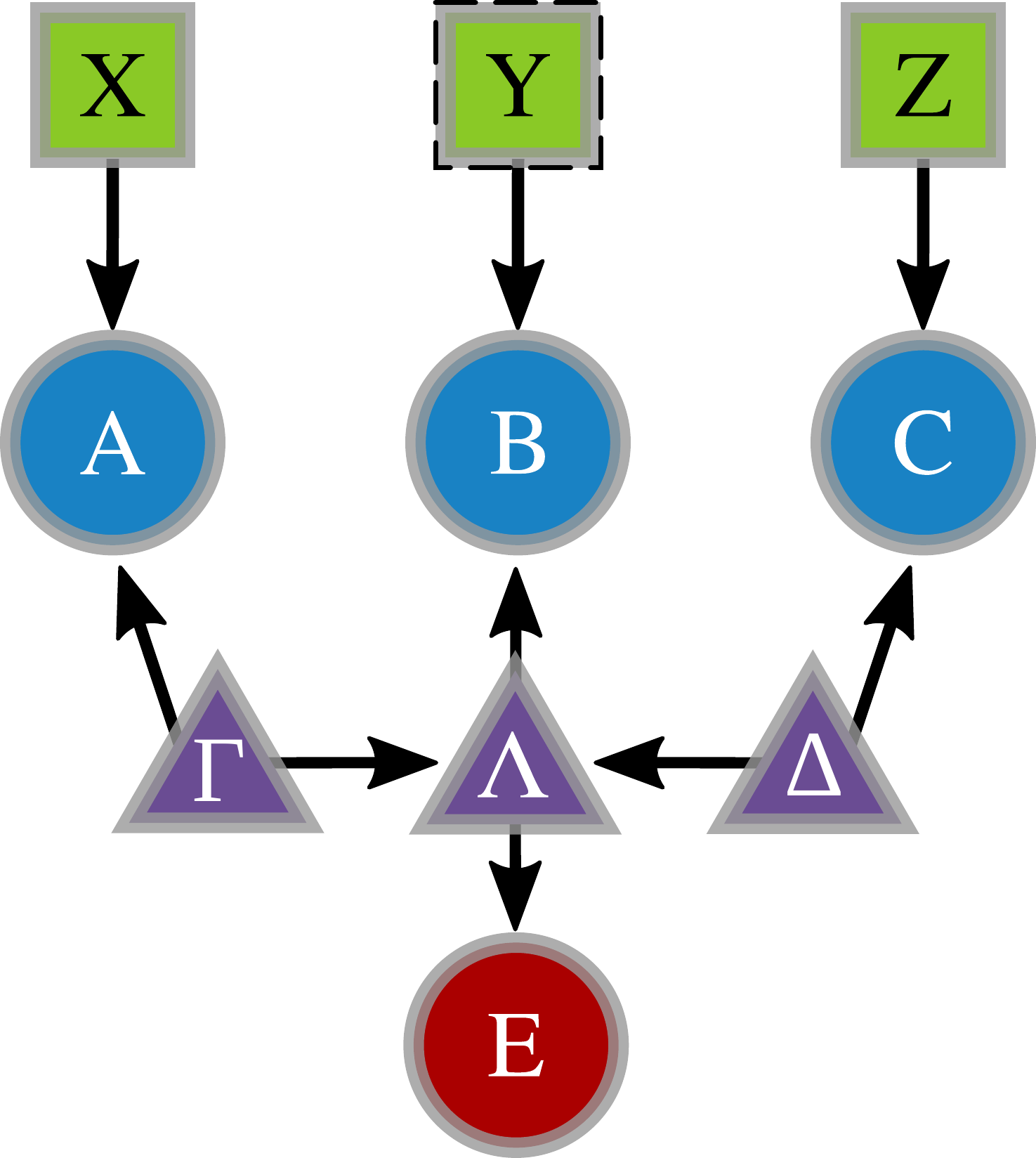}}}
\caption{\textbf{Different eavesdropping strategies within the bilocal scenario.} Eavesdropper actions are represented by red circles. \textbf{(a)} Double-Eavesdropper (DE) scenario reports a possible eavesdropping strategy within the bilocal scenario, accounting for the case of two distinct agents acting separately on the sources. \textbf{(b)} Weak-Eavesdropper (WE) scenario reports a single eavesdropper acting on both sources. \textbf{(c)} Strong-Eavesdropper (SE) scenario is equivalent to additionally supplying Eve with a further latent source. The dashed frame on the setting node $Y$ represents the possibility of performing both single- or multiple-setting measurements in the central node.}
\end{figure}

In contrast to a standard scenario with a single source, quantum networks introduce the constraint of independent sources, which allows multiple ways to model the eavesdropper's influence.
Specifically, in the bilocal scenario, different eavesdropping strategies can be considered, as illustrated in Fig.~\ref{fig:bilocal_dag_EF}. We assume that the eavesdropper can access only the sources locally, inheriting the limitations of the bilocal scenario.
Formally, this means that Eve can perform a POVM $E^e \otimes F^f$ on her share of the state $\rho_{A B_1 E_1} \otimes \rho_{A B_1 E_2}$, where $E^e$ and $F^f$ act only on the part $E_1$ and $E_2$ respectively.
We will refer to this as the ``Double-Eavesdropper'' (DE) scenario.
Note that we can also consider a more powerful eavesdropper that is allowed to measure a general POVM $E^e$ on both $E_1$ and $E_2$, as depicted in Fig.~\ref{fig:bilocal_dag_E}.
We will call this the ``Weak-Eavesdropper'' (WE) scenario.
Finally, we may also explore the worst-case scenario, where, while we retain the bilocal constraints for the nodes $A$ and $C$, we consider that Eve has full access to the central part of the network and can measure the same state as $B$.
We refer to this last scenario as the ``Strong-Eavesdropper'' (SE) scenario.
We represent this case by introducing an additional latent node $\Lambda$ affecting both $E$ and $B$ (see Fig.~\ref{fig:bilocal_dag_ΛE}). It is important to note that, while the WE and SE scenarios are equivalent when all variables are classical, in the quantum case, there can be a difference~\footnote{This is related to the known fact that the usual classical exogenization procedures do not work for quantum latent variables with incoming edges~\cite{wolfe2021quantum, centeno2024significance}.}. Since any eavesdropping strategy, including WE, can also be implemented in the SE case, the certified randomness in the latter scenario will always serve as a lower bound for the former. In the following analysis, we will concentrate mostly on the worst-case, i.e. the SE scenario, together with the DE scenario.

Analogously to Eq.(\ref{eqn:pguess}), one can define the global guessing probability in the bilocal scenario as
\begin{equation}
\label{eq:pguess_bilocal}
    G(ABC|E, xyz) = \sum_{a,b,c} p(abc,e=(abc)|xyz),
\end{equation}
which, again, represents the overall probability for an eavesdropper to correctly guess measurement outcomes.

In the first case, the information available to Eve can be bounded via the following optimization problem:
\begin{equation}
\label{eq:SDP_bilocal}
    \begin{aligned}
    \max_p \quad  & G(ABC|E, xyz) \\
    \text{s.t.} \quad &p(abce|xyz)  = \Tr (\rho_{ABCE} \cdot A_{a|x} \otimes B_{b|y} \otimes C_{c|z} \otimes E_{e}), \\
                      &p(abc|xyz)   = \sum_e p(abce|xyz) \\
                      &\rho_{ABCE}  = \rho_{AB_1E_1}\otimes\rho_{B_2CE_2}, 
    \end{aligned}
\end{equation}

Similarly, in the DE scenario, one can instantiate an analogous optimization problem with the crucial difference that the relevant guessing probability is now given by:
\begin{equation}
    G(ABC | EF, xyz) = \sum_{a,b,c} p(abc,e=(ab_0), f=(b_1c)|xyz), 
\end{equation}
where, as will be discussed below, $b_0$ and $b_1$ correspond to distinct bits associated to Bob's outcome. 
In both situations, one could also focus on the guessing probability corresponding only to the outcomes of the outer nodes, that is,
\begin{equation}
\label{eq:pguess_bilocal_red}
    G(AC | E, xz) = \sum_{a,c} p(ac,e=(ac)|xz).
\end{equation}
Its significance lies in the fact that the bilocal scenario can be seen as the prototype of a long-range quantum communication architecture, exploiting an intermediate node as a quantum repeater, exactly as it happens in event-ready Bell experiments~\cite{Zukowski1993,hensen2015loophole,Rosenfeld2017}.

\subsection{A numerical approach for randomness certification}
\label{sec:sdp}

\begin{figure*}[ht]
    \centering
    \subfloat[]{\label{fig:Hmin_teo_SE}{\includegraphics[width=0.5\textwidth]{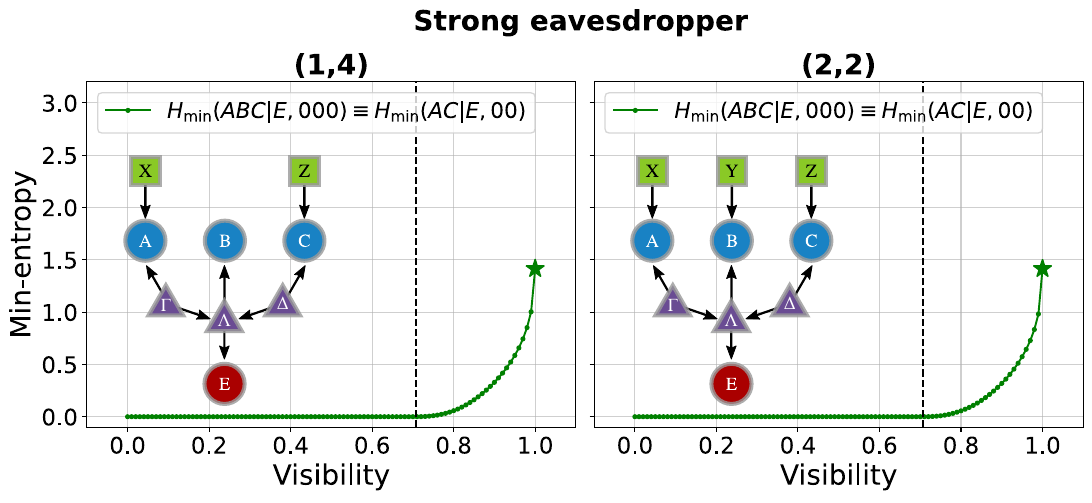}}} \hfill
    \subfloat[]{\label{fig:Hmin_teo_DE}{\includegraphics[width=0.5\textwidth]{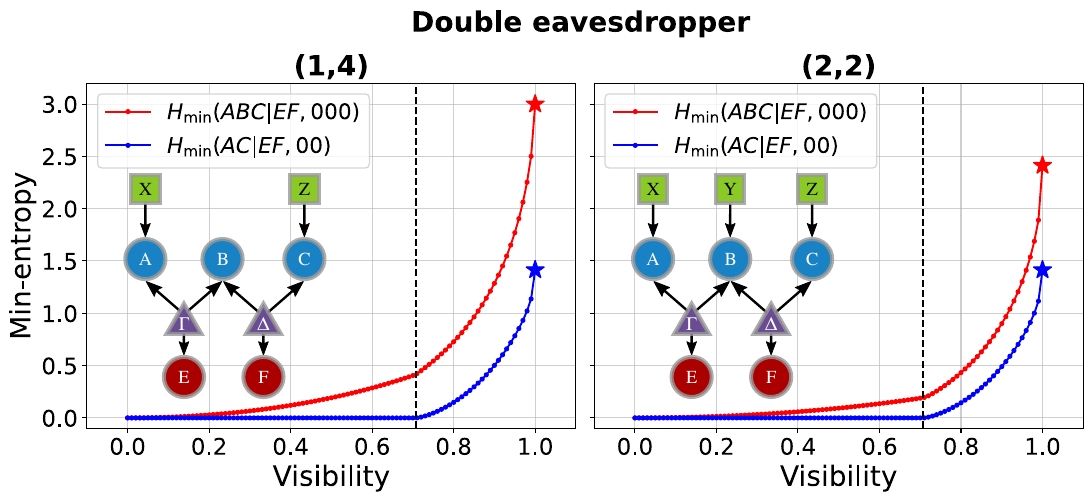}}}
    \caption{\textbf{Min-entropy for different configurations of the entanglement-swapping scenario.} Taking into account the possible eavesdropping strategies (SE or DE) and measurements performed in the central node ((1,4), (2,2)), we obtain four different configurations. For each of them, we report the min-entropy corresponding to the guessing probability obtained by solving the optimization problem in Eq.\ref{eq:SDP_bilocal} using the scalar extension technique. In particular, we plot the min-entropies associated either with the outer ($AC$) or all ($ABC$) parties, as a function of the visibility of the sources state. \textbf{(a)} In the strong eavesdropper scenario, both these quantities coincide and are jointly reported as green dots, while in the \textbf{(b)} double eavesdropper scenario, they are respectively illustrated as blue and red dots. The stars illustrate the theoretical upper bounds at unitary visibility (see Supplementary Information), which are saturated in every configuration of eavesdropping scenarios and measurement choices.
    The black dashed line shows the threshold visibility below which the states given by the sources, defined in  Eq.\eqref{eqn:werner_state}, can no longer violate the CHSH inequality.}
    \label{fig:num_results}
\end{figure*} 

To quantify the amount of certifiable randomness in the bilocal scenario, it is necessary to maximize the guessing probability of an eavesdropper. This probability is defined by the expressions in Eqs.\eqref{eq:pguess_bilocal} - \eqref{eq:pguess_bilocal_red}, subject to the constraint of observing a set quantum behavior described as in Eq.\eqref{eq:Qbehavior_bilocal}. The result of this optimization provides an estimate of the certifiable randomness in bits, quantified via the min-entropy, $H_{min} = -\log_2(G)$. However, in network scenarios, the independence of sources results in a non-convex set of correlations \cite{Chaves2016}, rendering standard techniques, such as the NPA hierarchy \cite{navascues2007bounding}, inapplicable. To address this challenge, the scalar extension technique \cite{pozas2019bounding} was developed. This method adapts the NPA hierarchy to account for the independence among the parties, enabling the optimization problem in Eq.\eqref{eq:SDP_bilocal} to be reformulated as a hierarchy of SDPs. Further details on the scalar extension method and its application to the bilocal scenario can be found in the Methods and Supplementary Information~\cite{supp}.

\begin{figure}[t!]
\centering
\includegraphics[width=0.99\linewidth]{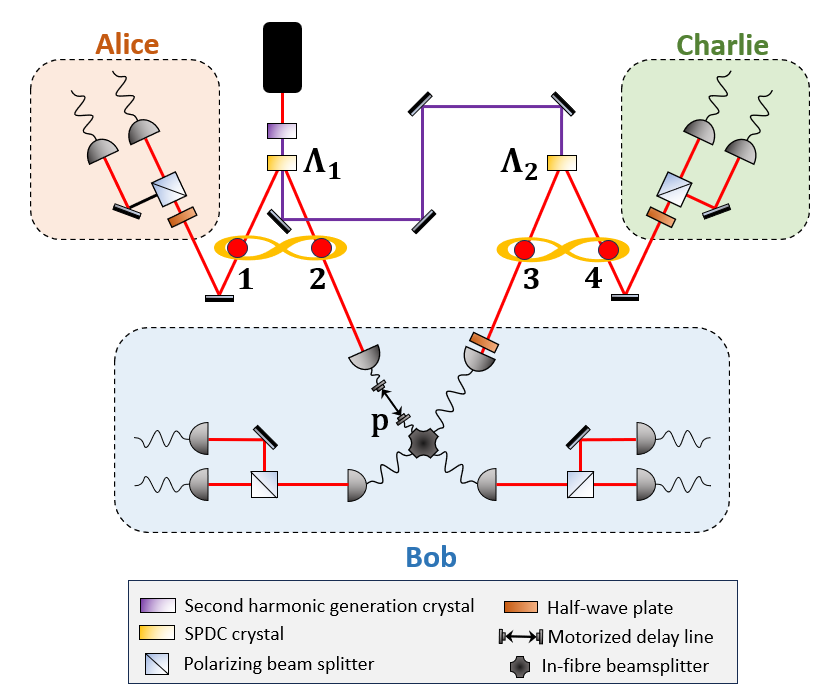}
\caption{\textbf{Experimental setup implementing the entanglement swapping network.} Two polarization-entangled photon pairs are generated via Spontaneous Parametric Down-Conversion (SPDC) in two separated non-linear crystals. Photons 2 and 3, one from each source, are directed to the central node Bob, while photon 1 (4) is directed to Alice (Charlie). The measurement performed in the central node is fixed and can either discriminate between $\ket{\Psi^-}$ and $\ket{\Psi^+}$, or between $\ket{\Phi^-}$ and $\ket{\Phi^+}$ depending on the configuration of the half-wave plate of Bob's station.}
\label{fig:exp_setup}
\end{figure}

To illustrate the general method, we start by considering the setup depicted in Fig.\ref{fig:exp_setup}. 
Each of the sources in the bilocal network is given by noisy quantum states modeled as
\begin{equation}
    \rho_{AB_1} = \rho_{B_2C} = v \dyad{\Psi^-}{\Psi^-} + (1-v)\frac{\mathbb{1}}{4},
\label{eqn:werner_state}
\end{equation}
where $v$ is visibility parameter \cite{werner1989quantum}. Concerning the measurement operators, two potential measurement strategies performed by the central node are considered: a single projective measurement on the Bell basis, or separable measurements given by $B_0 = \sigma_z \otimes \sigma_z$ and $B_1 = \sigma_x \otimes \sigma_x$. We will refer to these two choices using the labels ``(1,4)" and ``(2,2)", denoting the number of settings and outputs featured by Bob's measurements, respectively. In turn,  the outer node measurements have two possibilities, given by
\begin{equation}
\label{eq:AliceCharlie_meas}
    A_{0,1} = C_{0,1} = \frac{\sigma_z + (-1)^{(0,1)}\sigma_x}{\sqrt 2}.
\end{equation}

Taking these setups into account, we have solved the optimization problem in Eq.(\ref{eq:SDP_bilocal}), over the visibility range $v \in [0,1]$, as reported in Fig.\ref{fig:num_results} as well as in Tab.\ref{tab:biloc_results}.  
\\
\paragraph*{\textbf{Strong-Eavesdropper (SE) scenario.}} 
In the context of the strong-eavesdropper scenario for the measurement choices (1,4) and (2,2), we can certify up to $\approx 1.41$ bits of randomness when $v=1$. This value reaches its theoretical upper bound, as demonstrated by explicitly identifying a potential strategy for Eve. In this specific case of maximal visibility, the strategy involves a non-destructive Bell-state measurement of
the qubits directed to Bob, followed by a guess of Alice and Charlie's outcomes based on the expected probability distribution (see Supplementary Information). Moreover, Fig.\ref{fig:Hmin_teo_SE} shows that, in the SE scenario, it is possible to certify a non-zero amount of randomness as the visibility of the sources reaches the value $v = 1 / \sqrt{2}$, known to be the threshold above which a Werner state can violate the CHSH inequality.\\

\paragraph*{\textbf{Double-Eavestropper (DE) scenario.}}
Within this scenario, the threshold $v = 1 / \sqrt{2}$ is no longer valid since a non-zero amount of randomness can still be certified even for $v < 1/\sqrt{2}$. In addition, in this scenario, Eve can no longer perform projection measurements on the Bell basis, hence invalidating the previous optimal strategy. This is demonstrated in the numerical results shown in Fig.\ref{fig:Hmin_teo_DE}, where we achieve guessing probabilities as low as $G^{v=1}_{(2,2)}(ABC | EF, xz)=0.1875$ and $G^{v=1}_{(1,4)}(ABC | EF, xz)=0.125$, meaning that up to $\approx 2.41$ and $\approx 3$ bits of randomness can be certified for $v=1$ in the (2,2) and (1,4) measurement settings, respectively. Notably, under the assumption of independent eavesdroppers, a non-zero amount of certifiable randomness is observed across the entire range of visibilities in the scenario where the outcomes of all three nodes are guessed. While the randomness generated in this process originates from a combination of classical uncertainty and quantum correlations, this result might be valuable in practical scenarios where the assumption of eavesdropper independence is reasonable. It is worth highlighting that both results align with the intuitive observation that the independence of the eavesdroppers prevents them from collaboratively acting on the global system. This restriction inhibits the application of the Bell projection strategy on the central qubits, thereby limiting their predictive capabilities.

Special attention should be given to the certifiable randomness generated at the outer nodes, as this may represent the key figure of merit in long-distance communication scenarios where the central node functions solely as a repeater. Notably, the numerical results obtained, along with the theoretical upper bounds derived (see Supplementary Information), indicate that the amount of randomness reaches 1.41 bits for all combinations of measurement choices ((1,4) or (2,2)) and attack strategies, (SE or DE). This surpasses the typical value of 1.23 bits achieved through the violation of the CHSH inequality in a bipartite Bell scenario \cite{pironio2010random}.

\begin{figure*}[t!]
\centering
\hfill
\subfloat[]{\label{fig:Hmin_exp_SE}{\includegraphics[width=0.5\textwidth]{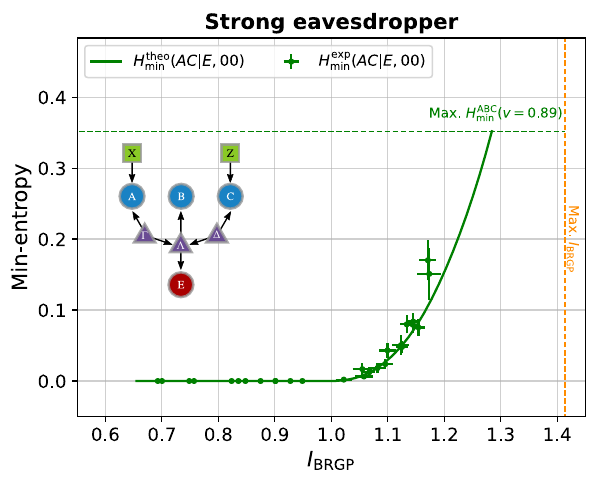}}} \hfill
\subfloat[]{\label{fig:Hmin_exp_DE}{\includegraphics[width=0.5\textwidth]{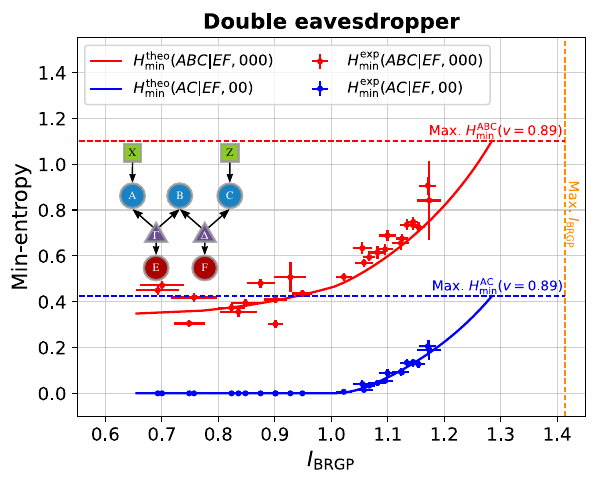}}}
\hfill
\caption{\textbf{Experimental min-entropy for the strong and double eavesdropper scenarios in the (1,4) measurement setup.} The min-entropy, derived from the guessing probability by solving Eq.(\ref{eq:SDP_bilocal}), is shown as a function of the violation of the bilocal inequality $I_{\mathrm{BRGP}}$. Theoretical predictions (solid curves) are compared with experimental data (crosses) for different values of $I_{\mathrm{BRGP}}$, controlled by adjusting the indistinguishability of the photons in the network's central node. \textbf{(a)} In the strong eavesdropper (SE) scenario, only the min-entropy of the outer nodes' outcomes are reported (green crosses and solid curve), as Bob's outcomes are fully known to the eavesdropper and do not contribute to the certifiable randomness. \textbf{(b)} In the double eavesdropper (DE) scenario, $H_{\mathrm{min}}(ABC|EF,000)$ (red) and $H_{\mathrm{min}}(AC|EF,00)$ (blue) differ and are shown as solid curves and crosses.  For both SE and DE cases, the maximum achievable min-entropy within the experimental visibility $v_{\mathrm{exp}}=0.89$ is indicated by dashed lines (green for $H_{\mathrm{min}}^{(\mathrm{SE})}(AC|E,000)$, red for $H_{\mathrm{min}}^{(\mathrm{DE})}(ABC|EF,000)$ and blue for $H_{\mathrm{min}}^{(\mathrm{DE})}(AC|EF,00)$), corresponding to the scenario where photons at Bob's station are perfectly indistinguishable. The orange dashed line represents the maximum violation of $I_{\mathrm{BRGP}}$.} 
\label{fig:sdp_vs_exp}
\end{figure*}

\subsection{Validation on experimental data}

To showcase a practical application of our method, we apply it to analyze the experimental data from Ref. \cite{Carvacho2017} that utilizes the photonic setup illustrated in Fig. \ref{fig:exp_setup} to provide the first randomness certification of non-local correlations within the bilocal scenario. In this setup, two non-linear crystals generate entangled photon pairs, serving as independent sources of quantum correlations. Alice's and Charlie's measurements are performed with polarization analyzers. Additionally, a partial Bell state measurement (BSM) is achieved through interference at an in-fiber beamsplitter, where a delay line adjusts the indistinguishability of the incoming photons.

\begin{table}[h!]
    \centering
   \begin{tabular}{c|cc|cc}
    & \multicolumn{2}{c|}{ \textbf{Strong}} & \multicolumn{2}{c}{\textbf{Double}}  \\
    & \multicolumn{2}{c|}{\textbf{eavesdropper}}  & \multicolumn{2}{c}{\textbf{eavesdropper}} \\
    ($N_{B}$, $|B|_{out}$) & \textbf{(1,4)} & \textbf{(2,2)}  & \textbf{(1,4)} & \textbf{(2,2)} \\[5pt]
     \hline \hline
    $H_{\mathrm{min}}(ABC|E(F),000)$  & 1.41 & 1.41 & 3.00 & 2.41 \\[5pt]
    $H_{\mathrm{min}}(AC|E(F),00)$ & 1.41 & 1.41 & 1.41 & 1.41 \\
    \hline
    \end{tabular} 
    \caption{\textbf{Min-entropy achieved with maximal visibility states:} Table accounting for the obtained numerical results. In particular, we report the min-entropies corresponding to states with unitary visibility for all the four configurations of the bilocal scenario that we considered: (1,4) and (2,2) indicate the number of settings and outputs in the central station (Bob).}
    \label{tab:biloc_results}
\end{table}

To compare the theoretical expectations and the experimental finding, we account for several sources of experimental imperfections: (I) the finite indistinguishability of photons 2 and 3, which directly impacts Bob's measurements; (II) an improved noise model that includes both white and colored noise in the quantum state; and (III) statistical fluctuations, which may cause the data to fall slightly outside the set of valid quantum behaviors. Further details on the experimental model are provided in the Supplementary Information. Additionally, we employed the NPA hierarchy, augmented with the scalar extension, to evaluate the certifiable randomness from the experimental data.\\

\paragraph*{\textbf{Strong-Eavesdropper scenario.}} In Fig. \ref{fig:sdp_vs_exp}, we compare the experimental and theoretical min-entropies as a function of the violation of the Branciard-Rosset-Gisin-Pironio bilocal inequality $I_{BRGP}$, as defined in ref.\cite{branciard_2012}, exhibiting excellent agreement. In the SE scenario, the experimental min-entropy on the outer nodes reaches $0.170 \pm 0.027$ bits, compared to its theoretical maximum of 0.35 bits, corresponding to the ideal case where Bob measures completely indistinguishable photons. In this scenario, we do not report the amount of randomness certifiable from all three nodes, as Bob's outcomes can always be predicted by an eavesdropper in the strong configuration, hence contributing zero bits to the min-entropy.

\paragraph*{\textbf{Double-Eavesdropper scenario.}}
In the context of the DE scenario, the experimental data allow us to certify up to $0.205 \pm 0.028$ random bits for external nodes $A$ and $C$ and up to $0.907 \pm 0.039$ random bits when including all three nodes, while the maximal theoretical predictions achieve 0.424 random bits (external nodes) and 1.10 random bits (all three nodes).\\
These results successfully validate our approach within a practical context and demonstrate that certifying a non-zero amount of secure randomness is feasible in a real-world network implementation.

\subsection{Tilted strategies for the Bilocal scenario}
\label{sec:tilted}
In the standard Bell scenario, the optimal strategies for randomness certification are not necessarily the ones that are maximally non-local~\cite{wooltorton2022tight, Woodhead2018randomnessversus}.
We are now going to consider similar strategies for the bilocal scenario, using different measurements in the $A$ and $C$ nodes, inspired by the tilted Bell inequalities, which are known to improve certified randomness in the Bell case~\cite{wooltorton2022tight}.
Specifically, we consider observables of the form:
\begin{align}
    \nonumber
    &A_0 = \sigma_z &A_1 = \cos \delta \sigma_x - \sin \delta \sigma_z\\
    &C_0 = \sigma_x &C_1 = \cos \delta \sigma_z - \sin \delta \sigma_x
\end{align}
while the central node $B$ performs the standard Bell state measurement as in the previous case.\\

\begin{figure}[t!]
\centering
\hfill
\subfloat[]{\label{fig:Hmin_tilted_SE}{\includegraphics[width=0.24\textwidth]{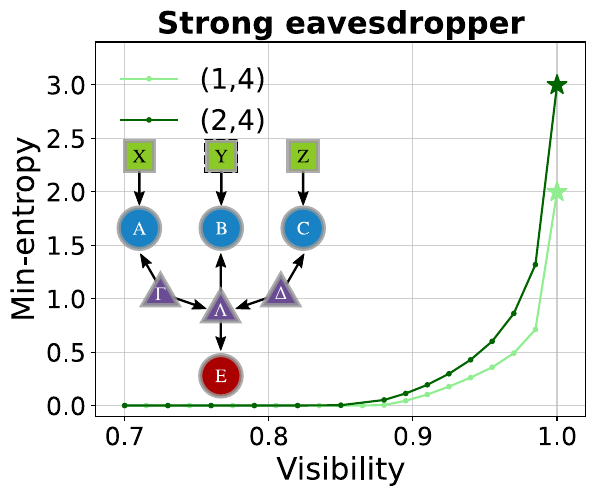}}} \hfill
\subfloat[]{\label{fig:Hmin_tilted_DE}{\includegraphics[width=0.24\textwidth]{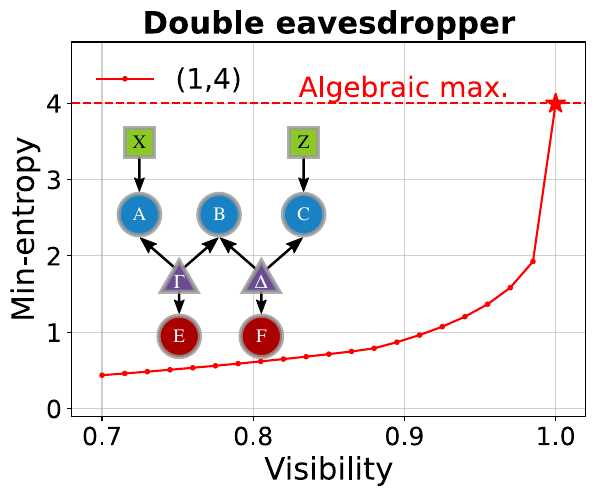}}}
\hfill
 \caption{\textbf{Min-entropy for alternative quantum strategies.}  analyze two quantum strategies using tilted Pauli operator: one with a single Bell state measurement (BSM) on B, denoted as (1,4) in the figure, and another with two measurement choices on B, one of which is a rotated BSM, (2,4) in the figure. \textbf{(a)} Min-entropy $H_{\min}(ABC|E)$ is shown for both strategies in the strong eavesdropper (SE) scenario, where the maximum values reach $3$ bits for the (2,4) case and $2$ bits for the $(1,4)$ case.
 \textbf{(b)} In the double eavesdropper (DE) scenario, as represented by the corresponding DAG, the (1,4) strategy allows reaching a maximum min-entropy of $4$ bits. The stars illustrate the maximum theoretical bound which are saturated. In particular, the min-entropy attained in the DE scenario reaches its algebraic maximum. This implies that the eavesdroppers do not have any information about the outcomes.} 
\label{fig:tilted_hmin}
\end{figure}

\paragraph*{\textbf{Strong-Eavesdropper scenario.}}
In this case, we find that it is possible to achieve the maximum of 2 bits per round for $\Hmin(AC|E)$ and the same value for $\Hmin(ABC|E)$ (see Fig.~\ref{fig:Hmin_tilted_SE}).
Similarly to the non-tilted case, this result can be explained by the fact that Eve can always guess the result of the BSM in the $B$ node, as described in the Supplementary Information~\cite{supp}. This suggests that the limit of 2 bits could be improved if we introduce a binary measurement setting $Y$ for the central node $B$.
Indeed, if we consider a protocol where $B_0$ is again the standard BSM, while $B_1$ projects on the rotated base:
\begin{align*}
\mathcal{B}_\theta = \{&\cos \theta \ket{00} + \sin \theta \ket{11}, \cos \theta \ket{01} + \sin \theta \ket{10},\\ 
&\sin \theta \ket{00} - \cos \theta \ket{11}, \sin \theta \ket{01} - \cos \theta \ket{10}\}
\end{align*}

we can get up to 3 bits of certified randomness as shown in Fig.~\ref{fig:Hmin_tilted_SE}.

\paragraph*{\textbf{Double- and Weak- Eavesdropper scenarios.}}
If, instead, we consider the DE and WE scenarios with the $(1,4)$ strategy, the restriction on using the same eavesdropping strategy dramatically increases the amount of certified randomness, as shown in Fig.~\ref{fig:Hmin_tilted_DE}.
In particular, it can be proved that, in the ideal case, we can certify up to $H_{\min}(ABC|E(F)) = 4$ for both the WE and DE scenarios (See Supplementary Information~\cite{supp}). In such a situation, the eavesdroppers have no information at all about the outcomes, and their best strategy is to uniformly guess them.

\section{Discussion}

The intrinsic randomness of quantum mechanics is fundamental for understanding the non-classical aspects of the theory. It has significant practical applications, including random number generation, randomness certification, and secure quantum communication. Although randomness in Bell-like scenarios --where a single source generates quantum correlations-- has been extensively studied and implemented experimentally, extending this framework to quantum networks with multiple independent sources remains largely uncharted. This challenge stems from the complexity of analyzing the non-convex set of correlations produced by independent sources \cite{Tavakoli_2022,Chaves2016}. We have addressed this gap by employing the scalar extension method \cite{pozas2019bounding}, which offers a reliable and robust approach to certify randomness within quantum networks.

To illustrate the power and versatility of our approach, we have focused on the entanglement-swapping network, a building block for quantum repeaters and an essential component in scalable quantum networks. This network enables different eavesdropping strategies, depending on whether Eve can access one or both entangled sources. In both scenarios, we demonstrated that up to $1.41$ bits of randomness can be certified between the network’s outer nodes, a value that surpasses the $1.23$ bits achievable through CHSH inequality violations between these nodes \cite{PhysRevLett.108.100402,Nieto-Silleras_2014}. This suggests that the source independence enforced by the network topology can offer a significant advantage in the randomness certification. When considering all the three network's nodes, we can exploit tilted measurement strategies to certify up to 4 bits of randomness, meaning that none of the outcomes can be known to potential eavesdroppers in such configuration. Additionally, we validated our approach by successfully quantifying the amount of randomness in the experimental data from the first photonic implementation of the bilocal network \cite{Carvacho2017}. 

Our findings and proposed methodology lay the groundwork for certification techniques in quantum networks of increasing size and complexity. They can also be applied to other network topologies that are attracting interest, such as the star network \cite{poderini2020experimental,wang2023certification,andreoli2017maximal}, triangle network \cite{polino2023experimental,wang2024experimental}, and the unrelated confounding scenario \cite{lauand2024quantum}. Furthermore, this approach could also find more sophisticated applications in networked quantum systems, including Bernoulli factory processes \cite{jiang2018quantum,liu2021general,hoch2024modular,rodari2024polarization} and blind quantum computation \cite{broadbent2009universal,polacchi2023multi,polacchi2024experimental}, contributing to the advancement of novel quantum communication architectures where randomness plays a central role.

\section*{Methods}

The numerical computation of the amount of randomness within the bilocal scenario is based on the scalar extension technique \cite{pozas2019bounding}, as the standard NPA hierarchy \cite{navascues2007bounding} cannot capture the causal independence relations that may arise among the network nodes due to the presence of independent sources. In the bilocal scenario, this is evident from the fact that the independence between Alice's and Charlie's nodes makes the corresponding probability distribution factorize as $\sum_b p(a,b,c|x,y,z) = p(a|x) p(c|z)$. Such an expression is non-linear and non-convex, so we can no longer characterize the quantum bilocal set of correlations using standard SDP relaxations.

In the standard NPA hierarchy, a \textit{moment matrix} of order $k$ is constructed as the matrix with entries $\Gamma_{ij} = \Tr (\rho O_i O_j)$, where $O_i(j)$ are products of the parties' measurement operators up to a length $k$. In the limit of $k \to \infty$, having $\Gamma \succeq 0$ certifies the membership of a given distribution to the set of quantum behaviors.

The main idea of scalar extension is to expand the set of operators that generate the moment matrix by incorporating additional elements derived from the products of actual operators and scalar terms, defined as the expectation values of operators (for example, terms such as $S_i \left \langle S_j \right \rangle$ or $S_i \left \langle S_j \right \rangle \left \langle S_k \right \rangle$). Such terms must be chosen so that the resulting extended moment matrix $\Tilde{\Gamma}$ has factorized entries that encode all the independence relations of the scenario of interest. Hence, linear expression in the extended moment matrix now suffices to express any independence among the parties, and optimization problems, such as maximization of the guessing probability over the set of bilocal quantum behaviors, can now be cast as SDPs using the scalar extension technique.

\section*{Data availability}
 The data that support the findings of this study are available in the Supplementary Information and from the corresponding author upon request.

\section*{Code availability}
 All the custom code developed for this study is available from the corresponding author upon request.

\section*{Acknowledgements} 
The authors acknowledge support from FARE Ricerca in Italia QU-DICE Grant n. R20TRHTSPA. G.C. acknowledges support from Sapienza Grant n. RG1241910DDF1480. RC acknowledges the Simons Foundation (Grant Number 1023171, RC), the Brazilian National Council for Scientific and Technological Development (CNPq, Grants No.307295/2020-6 and No.403181/2024-0), the Financiadora de Estudos e Projetos (grant 1699/24 IIF-FINEP) and the Otto Moensted Foundation visiting professorship. DP acknowledges funding from the MUR PRIN (Project 2022SW3RPY).

\section*{Competing interests}
The authors declare no competing interest.

\end{document}


\beginsupplement

\title{Supplementary Information: Experimental randomness certification in a quantum network with independent sources}

\author{Giorgio Minati}
\affiliation{Dipartimento di Fisica - Sapienza Universit\`{a} di Roma, P.le Aldo Moro 5, I-00185 Roma, Italy}
\author{Giovanni Rodari}
\affiliation{Dipartimento di Fisica - Sapienza Universit\`{a} di Roma, P.le Aldo Moro 5, I-00185 Roma, Italy}
\author{Emanuele Polino}
\affiliation{Dipartimento di Fisica - Sapienza Universit\`{a} di Roma, P.le Aldo Moro 5, I-00185 Roma, Italy}
\affiliation{Centre for Quantum Dynamics and Centre for Quantum Computation and Communication Technology Griffith University Yuggera Country Brisbane Queensland 4111 Australia}

\author{Francesco Andreoli}
\affiliation{ICFO - Institut de Ciències Fotòniques, The Barcelona Institute of Science and Technology, \\08860 Castelldefels, Spain }

\author{Davide Poderini}
\affiliation{International Institute of Physics, Federal University of Rio Grande do Norte, 59078-970, Natal, Brazil}
\affiliation{Università degli Studi di Pavia, Dipartimento di Fisica, QUIT Group, via Bassi 6, 27100 Pavia, Italy}

\author{Rafael Chaves}
\affiliation{International Institute of Physics, Federal University of Rio Grande do Norte, 59078-970, Natal, Brazil}
\affiliation{School of Science and Technology, Federal University of Rio Grande do Norte, 59078-970, Natal, Brazil}

\author{Gonzalo Carvacho}
\email[Corresponding author: ]{gonzalo.carvacho@uniroma1.it}
\affiliation{Dipartimento di Fisica - Sapienza Universit\`{a} di Roma, P.le Aldo Moro 5, I-00185 Roma, Italy}

\author{Fabio Sciarrino}
\affiliation{Dipartimento di Fisica - Sapienza Universit\`{a} di Roma, P.le Aldo Moro 5, I-00185 Roma, Italy}

\maketitle

\section{Noise modeling of the experimental apparatus}

When modeling the experimental distributions, we have to take into account two key aspects: the imperfect indistinguishability of the two incoming photons at Bob's measurement station and the presence of noise in the states generated by the sources. In the following, we will address each of these effects.

\subsection{Partial indistinguishability}
The partial indistinguishability directly affects the Bell State Measurement (BSM), since it mixes the detection of the Bell states $\ket{\Psi^-} \leftrightarrow \ket{\Psi^+}$ and $\ket{\Phi^-} \leftrightarrow \ket{\Phi^+}$, but it does not mix states belonging to different categories \cite{mattle1996dense}. We can model such an effect by substituting the projectors onto the Bell states with suitable effective POVMs:
\begin{equation}
\label{eq:effective_BSM}
    \begin{aligned}
    & \dyad{\Psi^-} \longrightarrow \hat{F}_1 = \frac{1+p}{2} \dyad{\Psi^-} + \frac{1-p}{2}\dyad{\Psi^+},\\
    & \dyad{\Psi^+} \longrightarrow \hat{F}_2 = \frac{1-p}{2} \dyad{\Psi^-} + \frac{1+p}{2}\dyad{\Psi^+},\\
    & \dyad{\Phi^-} \longrightarrow \hat{F}_3 = \frac{1+p}{2} \dyad{\Phi^-} + \frac{1-p}{2}\dyad{\Phi^+},\\
    & \dyad{\Phi^+} \longrightarrow \hat{F}_4 = \frac{1-p}{2} \dyad{\Phi^-} + \frac{1+p}{2}\dyad{\Phi^+},
    \end{aligned}
\end{equation}
where the parameter $p \in [0,1]$ quantifies the indistinguishability of the two photons. In particular, when $p=0$ the photons are distinguishable and the success rate of the measurements is 50\%, while in the case of indistinguishable photons ($p=1$) the effective POVMs  $\hat{F}_{1,2,3,4}$ coincide with the BSM. Experimentally, this parameter can be continuously tuned using a motorized delay line. 

\subsection{Noise in the SPDC sources of quantum states}
The quantum states generated through Spontaneous Parametric Down-Conversion (SPDC) sources, in our case, singlet states $\ket{\Psi^-}$, are affected by two distinct types of noise \cite{Cabello2005bell}: 
\begin{itemize}
    \item \textit{White noise:} corresponds to an isotropic depolarization of the state, thus mixing the singlet state with a completely mixed state:
    \begin{equation}
        \rho = v \dyad{\Psi^-} + (1-v) \frac{\mathds{1}}{4},
    \end{equation}
    where $\mathds{1}$ is the identity matrix and $v$ represents the visibility of the state.
\item \textit{Colored noise:} corresponds to a depolarization over a preferred direction, thus resulting in a statistical superposition of the singlet and a mix of $\ket{\Psi^-}$ and $\ket{\Psi^-}$:
\begin{equation}
    \rho = v \dyad{\Psi^-} + \frac{1-v}{2} \left( \dyad{\Psi^-} + \dyad{\Psi^+} \right).
\end{equation}
\end{itemize}
Therefore, we can model a state featuring both white and colored noise as:
\begin{equation}
\label{eq:exp_state}
    \rho_{v,c} = v \dyad{\Psi^-} + (1-v) \left[ \frac{1}{2} c \left( \dyad{\Psi^-} + \dyad{\Psi^+} \right) + (1-c) \frac{\mathds{1}}{4} \right],
\end{equation}
where $v$ is the overall visibility of the state, while $c$ represents the fraction of colored noise over the total noise.

\subsection{Randomness estimation in presence of experimental noise}

To understand how the amount of certified randomness can be affected by the experimental imperfections, we solved the guessing probability maximization problem for different regimes of noise. In particular, we considered the bilocality scenario in the case in which two eavesdroppers separately act on the sources and Bob is carrying out Bell state measurements. Hence, the numerical evaluation of the guessing probability in the presence of experimental noise consists of solving the following optimization problem:
\begin{equation}
\label{eq:SDP_bilocal_expnoise}
    \begin{split}
    \text{max} \quad  & G(A(B)C|EF, x(y)z) \quad \text{s.t.} \\
    p(abc|xyz) & = \Tr (\rho_{ABC}^{\mathrm{exp}} \cdot A_{a|x} \otimes B_{b|y}^{\mathrm{exp}} \otimes C_{c|z}), \\
    p(abc|xyz) &  = \sum_e p(abc,e|xyz),
    \end{split}
\end{equation}
where $\rho_{ABC}^{\mathrm{exp}} = \rho_{AB}^{v,c} \otimes \rho_{BC}^{v,c}$ with $\rho^{v,c}$ given by the noisy model of the experimentally generated quantum states reported in Eq.\ref{eq:exp_state}, while $B_{b|y}^{\mathrm{exp}}$ corresponds to the effective Bell state measurements described in Eq.\ref{eq:effective_BSM}.

\begin{figure}[H]
    \centering
    \includegraphics[width=1\linewidth]{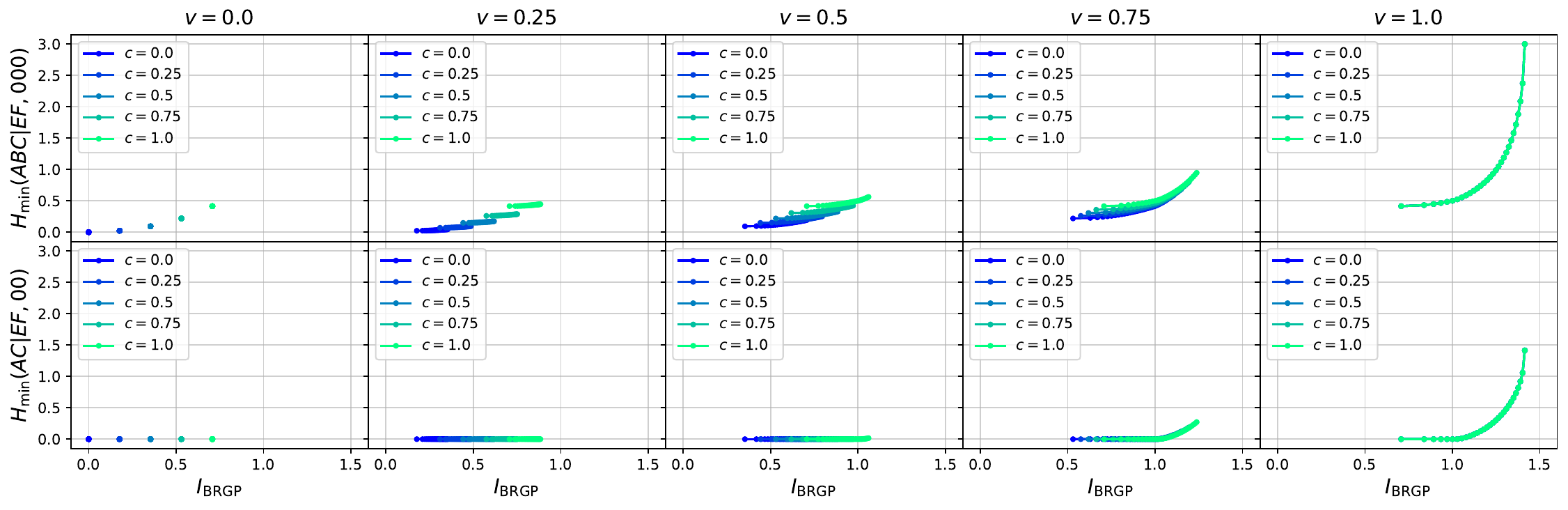}
    \caption{\textbf{Min-entropy in presence of experimental noise.} Here, we report the numerically computed min-entropy (solving the optimization problem in Eq.(\ref{eq:SDP_bilocal_expnoise})) for different regimes of noise. In particular, we show the behavior of both $H_{\mathrm{min}}(ABC|EF,000)$ (upper panels) and $H_{\mathrm{min}}(AC|EF,00)$ (lower panels) as a function of the bilocal inequality $I_{\mathrm{BRGP}}$ when we vary the indistinguishability parameter $p$, effectively changing Bob's measurements (see Eq.(\ref{eq:effective_BSM})). Moreover, in each panel of the figure we consider different values of the visibility $v$, comparing different plots corresponding to different amounts of colored noise in the quantum states.} 
    \label{fig:Hmin_exp_model}
\end{figure}
In Supplementary Figure \ref{fig:Hmin_exp_model}, each plot illustrates how the min-entropy evolves as the states exhibit varying values of visibility $v$ and a fraction of colored noise $c$, while systematically varying the indistinguishability parameter across its full range, $p \in [0,1]$. The resulting min-entropy has been reported as a function of the violation of the bilocal inequality $I_{BRGP}$. Moreover, this analysis has been done for both cases: when the eavesdropper attempts to guess, either the outcomes of all three parties (upper panels of Supplementary Figure \ref{fig:Hmin_exp_model}) or only the outer ones (lower panels of Supplementary Figure \ref{fig:Hmin_exp_model}).

\section{Numerical technique: scalar extension in the bilocal network}
The scalar extension method \cite{pozas2019bounding} aims to allow NPA-like relaxations in network scenarios, i.e. to approximately characterize the network quantum set, a task which is not trivial due to the non-convexity of this set.
The difficulty in employing the NPA method in networks stems from one of the characterizing features of such scenarios, i.e. the presence of multiple independent sources. 
\begin{figure}[H]
    \centering
    \includegraphics[width=0.8\linewidth]{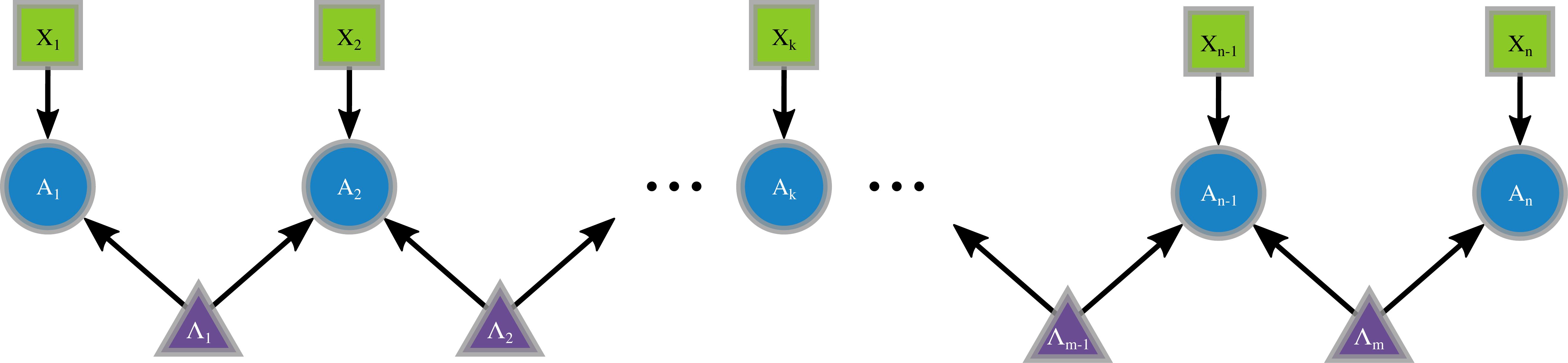}
    \caption{\textbf{DAG of the chain network scenario.} In this scenario, $m$ bipartite sources distribute correlations to pairs of the $n=m+1$ parties, which perform local measurements choosing among settings described by the variables $\{ X_1, \ldots, X_n \}$ and obtaining the outcomes $\{ A_1, \ldots, A_n \}$.} 
    \label{fig:line_scenario}
\end{figure}
We can consider, for simplicity, the case of a chain network scenario, whose DAG is depicted in Supplementary Figure \ref{fig:line_scenario}. It is possible to show that behaviors arising from such structure fulfill the following condition:
\begin{equation}
    \label{eq:example_factorization}
\begin{aligned}
    & \sum_{a_k} p(a_1, \ldots, a_{k}, \ldots, a_n|x_1, \ldots, x_k, \ldots, x_n) =\\ & = p(a_1, \ldots, a_{k-1}|x_1, \ldots, x_{k-1}) p(a_{k+1}, \ldots, a_n| x_{k+1}, \ldots, x_n),
\end{aligned}
\end{equation}
where $(a_1, \ldots,a_n)$ and $(x_1,\ldots,x_n)$ respectively denote the outcomes and the setting of the measurements performed by each party.\\
When we marginalize one of the non-extremal parties, the distribution factorizes, meaning that the corresponding parties are conditionally independent.\\
In the context of causal modeling, the independence relations can be identified by resorting to the notion of d-separation. In this case, we can say that when the path connecting two nodes contains a structure with two converging arrows (called \textit{collider}), then the corresponding variables are conditionally independent. Therefore, observing the DAG depicted in Supplementary Figure \ref{fig:line_scenario}, it is possible to recognize that the collider with the middle node in $A_k$ lies in the path between the groups of outcomes $(A_1,\ldots,A_{k-1})$ and $(A_{k+1}, \ldots, A_n)$. Then, we deduce that the factorization reported in Eq.\ref{eq:example_factorization} encodes the conditional independence $((A_1,\ldots, A_{k-1}) (A_{k+1},\ldots,A_n)|A_k)$).\\
For this reason, the inability to apply the NPA method in network scenarios can be attributed to the causal separations that emerge in the presence of independent latent variables. Expressions akin to the factorized distribution in Eq.\ref{eq:example_factorization} are both nonlinear and non-convex, features which make it impossible to cast the characterization of the network quantum set into an SDP problem. The aim of scalar extension is indeed to overcome this obstacle by encoding the independence relations arising from the network structure in constraints which are linear in the entries of the moment matrix, then compatible with the SDP relaxations exploited by the NPA method.

\subsection{Construction of the Method}
The \textit{scalar extension} main idea is, indeed, an extension of the moment matrix employed in the standard NPA hierarchy. In particular, we complement the set of operators $\mathcal{O}$ that generate the matrix $\mathbf{\Gamma}$ with additional operators composed of the product of operators multiplied by the expected value of other products of the operator. Then, we will extend the set $\mathcal{O}$ with operators of the form $S_i \langle S_j \rangle$ or $S_i \langle S_j\rangle \langle S_k \rangle$. Once we add such operators, there will be factorized quantities among the new entries in the moment matrix $\Tilde{\mathbf{\Gamma}}$ which allows us to set up linear relations. It is fundamental to choose the extension variable to make the factorized entries encode all independencies featured in a given scenario. At this stage, the NPA method can be applied in the usual manner by constructing a matrix $\Tilde{\mathbf{\Gamma}}$, where certain entries are fixed by the observed distribution, while those corresponding to unobservable measurements are treated as variables. These variables are then optimized through a semidefinite programming (SDP) problem to verify the existence of a matrix satisfying $\Tilde{\mathbf{\Gamma}} \succeq 0$. If the solution is positive, it confirms that the observed distribution belongs to the network quantum set.\\
Moreover, by construction, the existence of $\Tilde{\mathbf{\Gamma}} \succeq 0$ implies the existence of $\mathbf{\Gamma} \succeq 0$, since the latter is a principal submatrix of the former. Instead, if we cannot find any scalar extension certificate, it means that the proposed causal explanation that we proposed is not compatible with the observed distribution. If, in addition, any matrix $\mathbf{\Gamma}$ can be found, we can also conclude that the distribution is incompatible with any measurements on a global quantum state.\\
To better understand the notions about the scalar extension method that we have introduced so far, we can consider the bilocal scenario as an exemplary instance of the network. Employing the d-separation criterion, we deduce that Alice's and Charlie's nodes are d-separated by Bob's node. As a consequence, all the entries of $\Tilde{\mathbf{\Gamma}}$, which only contains Alice's and Charlie's operators must factorize. For instance, we can consider the moment matrix generated by the set of operators $\mathcal{O} = \{ \mathds{1}, A_{0|0}A_{0|1}, C_{0|0}C_{0|1}, \langle A_{0|0}A_{0|1} \rangle \mathds{1}\}$.
\begin{equation}
\Tilde{\mathbf{\Gamma}}=
    \begin{blockarray}{ccccc}
    &\mathds{1} & A_0A_1  & C_0C_1 & \langle A_0A_1 \rangle \mathds{1} \\
        \begin{block}{c(cccc)}
          \mathds{1}                          & 1 & v_1 & v_2 & v_3 \\
          (A_0A_1)^{\dag}                     &   &  1  & v_4 & v_5 \\
          (C_0C_1)^{\dag}                     &   &     &  1  & v_6 \\
          \langle A_0A_1 \rangle^* \mathds{1} &   &     &     & v_7 \\
        \end{block}
    \end{blockarray}
\end{equation}
where we reported only the upper triangular matrix, since $\Tilde{\mathbf{\Gamma}}$ is Hermitian. A priori, we should optimize over all the variable $v_i$ to seek the values that make the moment matrix positive semidefinite. However, the presence of the extra operator $\langle A_0A_1 \rangle \mathds{1}$ produces a series of relationships among the variables, making some dependent on others, for instance, we can deduce $v_1=v_3$, $v_5=v_7$, and $v_4=v_6^*$. In particular, the latter of these equalities is crucial from a causal perspective, since it imposes the factorization $ \langle A_0A_1 C_0C_1\rangle =  \langle A_0A_1 \rangle  \langle C_0C_1 \rangle$ i.e. the independence constraint which arises from the network structure of the bilocal scenario. The same causal independence could be imposed by constraints like $v_4 = v_1^* v_2$ and $v_5 = |v_1|^2$. However, due to their nonlinearity, these relationships cannot be directly incorporated into an SDP problem. Consequently, we will not enforce them.\\

\section{Experimental data pre-processing}
The experimental data obtained in \cite{carvacho2017experimental} consists of the coincidence counts $N(abc|xz)$ measured in the single-photon detectors for possible measurement outcomes and settings. Given the statistical noise affecting the experimental counts, it is possible that the reconstructed probability distribution $p_{\mathrm{exp}} (abc|xz) = N(abc|xz) / \sum_{a,b,c} N(abc|xz)$ may violate the so-called no-signaling (NS) constraints, imposing the impossibility of signaling among space-like separated parties. As the quantum set of correlations is a subset of the NS one, when this occurrence takes place, an SDP optimization aimed at maximizing the guessing probability would not be feasible if constrained to such a distribution. To overcome this problem, we perform a "projection" of the experimentally reconstructed distributions over the NS set of correlations through the following maximum likelihood problem:
\begin{equation}
\label{eq:NS_proj}
    \begin{split}
    \text{max} \quad   \log \mathcal{L}  & \equiv \log \left(\prod_{a,b,c,x,z} p(abc|xz)^{N(abc|xz)} \right) \quad \text{s.t.} \\
    \sum_{a,b,c} p(abc|xyz)  & = 1, \\
    \sum_a p(abc|xyz) & = \sum_a p(abc|x' yz) \quad \forall x, x' \\
    \sum_c p(abc|xyz) &  = \sum_c p(abc|xyz') \quad\forall z, z',\\
     p(ac|xz) &= p(a|z)p(c|z),
    \end{split}
\end{equation}
where the first constraint normalizes the variable $p(abc|xz)$, the second and the third ones impose the NS constraints, while the fourth establishes the conditional independence among the outer nodes outcomes $a$ and $c$.\\
The solution to this problem provides the closest distribution to $p_{\mathrm{exp}} (abc|xz)$ that satisfies the NS constraints, which, therefore, can be safely employed to evaluate the amount of certifiable randomness from a given experimental distribution.

\section{Lower bounds to the guessing probability}

In this section, we derive some lower bounds to the $P_{\mathrm{guess}}$.
This approach is complementary to the numerical results based on the NPA hierarchy and the scalar extension. 
While the latter provides a lower bound to the certifiable bits, depending on the chosen hierarchical order, here we define some minimal methods to identify a \emph{possible} physical strategy for Eve in a generic network. 
This strategy can differ from the optimal one, thus providing an upper bound to the certifiable bits in a given network scenario.

\subsection{Eavesdropping strategies in general networks}
Consider a generic network defined by a DAG as in Fig.~1 in the main text.
Each DAG is composed of three different kinds of nodes: the \emph{latent} nodes, representing the network sources, the \emph{eavesdropper} nodes, and the \emph{observables} nodes, with (optionally) the associated \emph{setting} node.
To each latent variable, we associate a quantum state described by the density operator $\rho_\Lambda \in \mathcal{L}(\mathcal{H}_\Lambda)$.
We call $\rho = \bigotimes_{\Lambda} \rho_\Lambda \in \mathcal{L}(\mathcal{H})$ the overall quantum state produced by the sources, where $\mathcal{H}$ is the corresponding global Hilbert space.
To each observed (or eavesdropper) variable $A$ we associate a POVM measurement described by the operators $\left\{\mathcal A_{x}^{a}\right\}_a$, optionally dependent on a setting $x$. 
Measurement operators relative to different nodes should be pairwise commuting.
The distribution of the outcomes of the observable and eavesdropper nodes is then given by:
\begin{equation}
    P(a_1,\ldots,a_n, e_1, \ldots, e_m|x_1,\ldots,x_n;\rho) = 
    \tr \left(\mathcal A_{x_1}^{a_1}\cdots {\mathcal A}_{x_n}^{a_n} {\mathcal  E}^{e_1} \cdots {\mathcal E}^{e_m} \; \rho \right)
\end{equation}
where the directed edges determine on which part of the space $\mathcal{H}$ of the latent nodes, the measurements ${\mathcal A}_{x}^{a}$ and ${\mathcal E}^{e}$ act non-trivially.

To simplify the notation we will write $\mathcal A_{\vec x}^{\vec a} = \prod_{i} \mathcal A_{x_i}^{a_i}$ for the measurement operator on all observable nodes with settings $\vec x$ and outcomes $\vec a$.

The goal of the eavesdropper is to guess the outcome of the observable nodes for some specific values of the settings (e.g. $\vec x=0$ without loss of generality) with the most efficient strategy available.
This corresponds to maximizing the \emph{guessing probability} given by:
\begin{equation}
    P_{\rm guess}(\mathcal A_0 ) \equiv \Sum_{\vec a,\vec b}P(\vec e=\vec a, \vec a,\vec b|\vec x=0; \rho).
\end{equation}
where the variables $\vec b$ represent potential bits that influence the distribution, but which Eve is not interested in guessing. 
These would be, for example, Bob's outcomes in the case labeled AC of the main text, which corresponds to a bilocality scenario where Eve only aims to guess the bits produced by Alice and Charlie.
Conversely, the certification measurements are meant to detect the action of an eavesdropper: to avoid being noticed, the eavesdropper must guarantee that:
\begin{equation}
    \Sum_{\vec e} P(\vec e,\vec a ,\vec b|\vec x, \vec y; \rho) = P_Q(\vec a,\vec b |\vec x, \vec y), \quad \forall \vec a,\vec b,\vec x,\vec y
\end{equation}
where $\vec x, \vec y$ are the settings associated to the parties $A$ and $B$ respectively and $P_Q(\vec a ,\vec b| \vec x, \vec y) = \tr(\rho \cdot \mathcal A_{\vec x}^{\vec a} \mathcal B_{\vec y}^{\vec b})$ is the probability distribution of the measurements without the intervention of the eavesdropper.

In the following, we describe some generic strategies in which Eve exploits some protocol vulnerabilities. 

\begin{itemize}
    \item \textbf{Uniform guess.} 
    The most trivial strategy available to Eve is uniformly guessing an outcome. This provides the lowest bound to the guessing probability, i.e. the utmost bound to the certifiable randomness. After an outcome $\vec a$ is extracted, the conditioned probability of correctly guessing $\vec a$ using this basic strategy is 
    \begin{equation}
        P(e=\vec a|\vec a,\vec b, 0; \rho) = \dfrac{1}{N_{\mathcal A}},   
    \end{equation}
    where $N_{\mathcal A}$ is the number of possible outcomes of the extraction measurement $\mathcal A_0$. This gives the trivial bound $P_{\rm guess}(\mathcal A_0 ) \geq P_{\rm uniform}(\mathcal A_0 ) = N_{\mathcal A}^{-1}$ for the guessing probability.
    %
    In the dichotomic case, one has that $N_{\mathcal A} = 2^N$, leading to $H_{\rm min} \leq N$, where $N$ is the number of observable nodes. 
    \item \textbf{Informed guess.} 
    Eve can choose to guess the outcomes of $\mathcal A_{0}$ based on the knowledge of the expected quantum distribution of the extraction outcomes. Specifically, Eve should bet on the most probable result $\vec{a^*}$ for each value of $\vec b$. 
    This leads to
    \begin{equation}
        P(e=\vec a|\vec b,0;\rho) = \displaystyle \max_{\vec a} P_Q(\vec a|\vec b,0;\rho) = P_Q(\vec{a^*}|\vec b,0;\rho).
    \end{equation}
    We call this method ``informed guess", which leads to the overall guessing probability
    \begin{equation}
        \label{eq:intro_info_guess_basic}
        P_{\rm guess}(\mathcal A_0 ) \geq P_{\rm info} (\mathcal A_0; \rho) = 
        \Sum_{\vec b} P(\vec e=\vec a|\vec b,0;\rho) P_Q(\vec b|0;\rho) =
        \Sum_{\vec b}P_Q(\vec{a^*}|\vec b,0;\rho) P_{Q}(\vec b|0;\rho).
    \end{equation}
    If these outcomes are uncorrelated (i.e. we have a uniform distribution $P_Q(\vec a|\vec b,0;\rho)=1/N_{\mathcal A}$), then the guessing probability becomes the lowest possible $P_{\rm info} (\mathcal A_0;\rho) = 1/{N_{\mathcal A}}$. 
    On the contrary, the more correlated the outcomes are, the more Eve can guess correctly using this basic strategy. When Eve attempts to guess all the bits produced in the nodes, then no variable $\vec b$ is present, and the informed guess probability reduces to $P_{\rm info}(\mathcal A_0;\rho) = \max_{\vec a} P_Q(\vec a|0;\rho)$.
    
    \item \textbf{Node vulnerability.} 
    A more sophisticated strategy consists of identifying some vulnerabilities in the protocol, which allows to perform a projective measurement $\mathcal E^{\vec e} = \mathbb P_{\vec e}$ (with distinct outcomes $\vec e$) that cannot be detected by the nodes. 
    
    Then, Eve can either exploit the information gained by the knowledge of $\xi$ or the correlations introduced by projecting the state, to define the guess probability 
    \begin{equation}
        \label{eq:intro_info_guess_advanced}
        P_{\rm guess}(\mathcal A_0 )\geq P_{\rm info}(\mathcal A_0 ) = \Sum_\xi \tr(\rho \mathbb P_{\vec e}) P_{\rm info} (\mathcal A_0; \rho_{\vec e}),\quad
        \rho_{\vec e} = \frac{\mathbb P_{\vec e} \rho \mathbb P_{\vec e}}{\tr(\rho \mathbb P_{\vec e})}.
    \end{equation}
    A specific example of such a mechanism can be identified when all the measurements of a node $j$ commute, i.e. $[\mathcal A_{x}^{a_j},\mathcal A_{x'}^{a_j}]=0 \quad \forall x ,x'$. 
    In this case, then Eve can define $\mathcal E^{e_j}$ as the projection of the initial state onto the shared eigenbasis of the operators $\{\mathcal A_{x}^{a_j}\}_x$. 
    In case of degeneracies, then the proper unitaries must be found that define the shared eigenbasis. 
    Using this approach, Eve will be able to distinguish all the outcomes, allowing her to guess with certainty the outcome of that node. 
    This is trivially possible when a node only performs a single measurement, as in the bilocal case where the central node Bob always performs a Bell-state projection.
\end{itemize}

\section{Bilocality scenario}
\label{sec:min_model_network}
Here, we discuss how to apply the strategies defined above to the case of a bilocality scenario. 
In the standard strategy used in this scenario, the state is generated by two sources and reads
\begin{equation}
    \rho  = \ket{\Psi^-_{AB_1}}\!\bra{\Psi^-_{AB_1}} \otimes \ket{\Psi^-_{B_2C}}\!\bra{\Psi^-_{B_2C}},
\end{equation}
where we focus on the case of a pure state (absence of noise). 
The external nodes (Alice and Charlie) perform the measurements
\begin{equation}
    \mathcal A_0=\mathcal C_0 = \frac{\sigma_z+\sigma_x}{\sqrt{2}},\quad
    \mathcal A_1=\mathcal C_1 = \frac{\sigma_z-\sigma_x}{\sqrt{2}}.
\end{equation}

We consider the two possible scenarios, where the central node either performs only a Bell-state measurement with four outcomes
\begin{equation}
    \mathcal B^{(14)} = 
      b_1\ket{\Psi^+}\bra{\Psi^+}
    + b_2\ket{\Psi^-}\bra{\Psi^-}
    + b_3\ket{\Phi^+}\bra{\Phi^+}
    + b_4\ket{\Phi^-}\bra{\Phi^-},
\end{equation}
or can choose between two possible, two-outcome measurements
\begin{equation}
    \mathcal B^{(22)}_y = (1-y)\sigma_z\otimes \sigma_z+y\sigma_x\otimes \sigma_x,\quad y=0,1.
\end{equation}

\subsection{Minimal strategy for Eve}
\label{sec:min_strategy}
Eve can apply some minimal strategies without performing any measurements. The ``uniform guess" bounds the maximum number of certifiable bits to be
\begin{align}
\nonumber
    &H_{\rm min}^{(14)}(\mathcal A_0,\mathcal B^{(14)},\mathcal C_0)\leq 4,
    &&H_{\rm min}^{(22)}(\mathcal A_0,\mathcal B^{(22)}_0,\mathcal C_0)\leq 3,\\
    &H_{\rm min}^{(14)}(\mathcal A_0,   \mathcal C_0)\leq 2,
    &&H_{\rm min}^{(22)}(\mathcal A_0,  \mathcal C_0)\leq 2.
\end{align}
A tighter bound can be obtained by the ``informed guess" strategy without any additional operation (i.e. \eqref{eq:intro_info_guess_basic}). Considering the bilocality measurements, one obtains an improved estimation
\begin{align}
\nonumber
    &H_{\rm min}^{(14)}(\mathcal A_0,\mathcal B^{(14)},\mathcal C_0)\lesssim 3 ,
    &&H_{\rm min}^{(22)}(\mathcal A_0,\mathcal B^{(22)}_0,\mathcal C_0)\lesssim 2.41,\\
    &H_{\rm min}^{(14)}(\mathcal A_0,  \mathcal C_0)\lesssim 1.41,
    &&H_{\rm min}^{(22)}(\mathcal A_0,   \mathcal C_0)\lesssim 1.41.
\label{eq:bilo_info_guess_basic_bounds}
\end{align}

Hereafter, we discuss how an improved bound can be obtained by studying the vulnerabilities of the bilocality protocol, before using an ``informed guess" strategy, as described in Eq.~\eqref{eq:intro_info_guess_advanced}. This is specifically relevant when considering the causal structure described in Fig.1-c of the main text. 

\begin{figure}[h]
\centering
\subfloat[]{\label{fig:bilocal_dag_EF}{\includegraphics[width=0.2\textwidth]{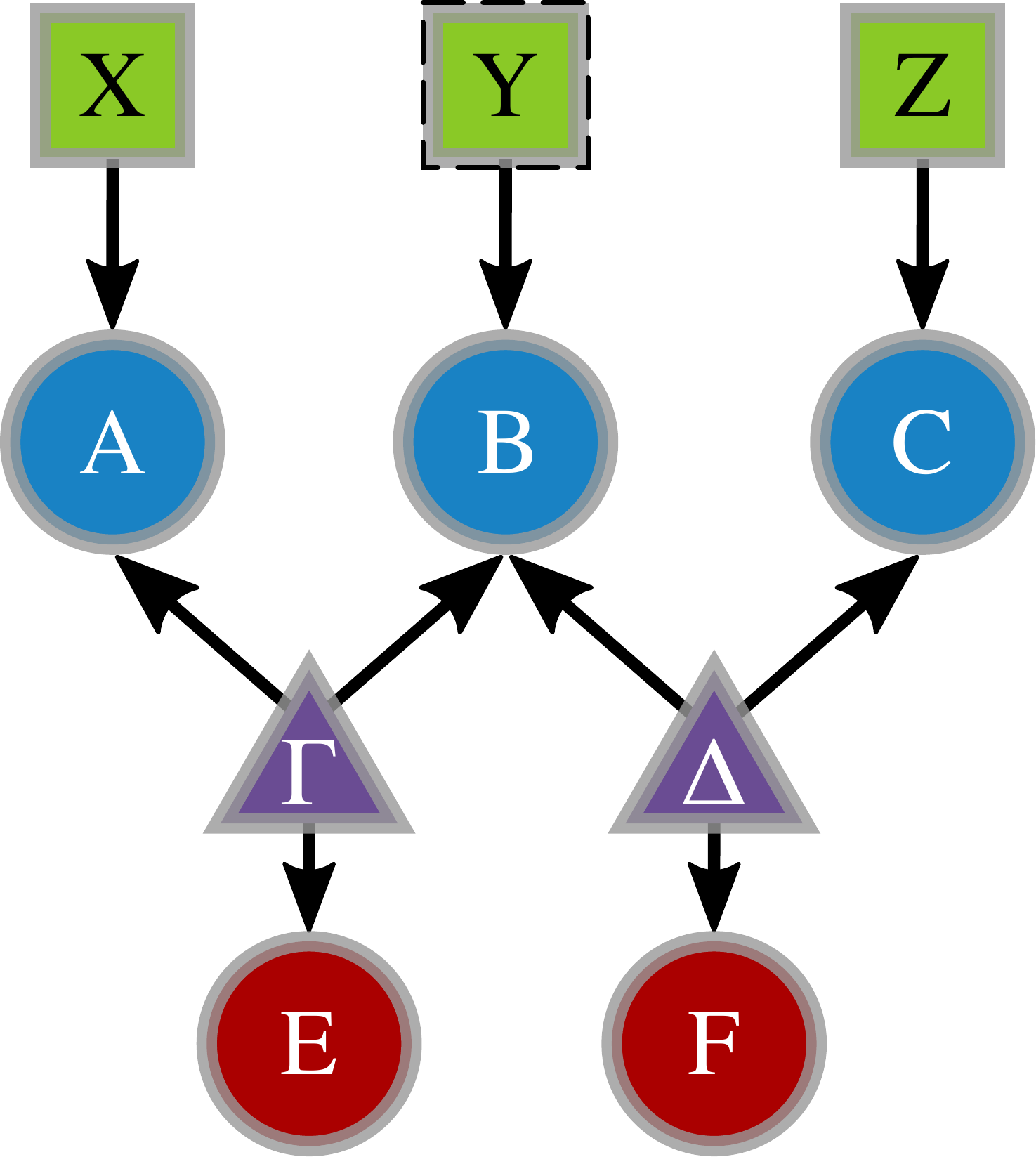}}}
\hspace{2cm}
\subfloat[]{\label{fig:bilocal_dag_E}{\includegraphics[width=0.2\textwidth]{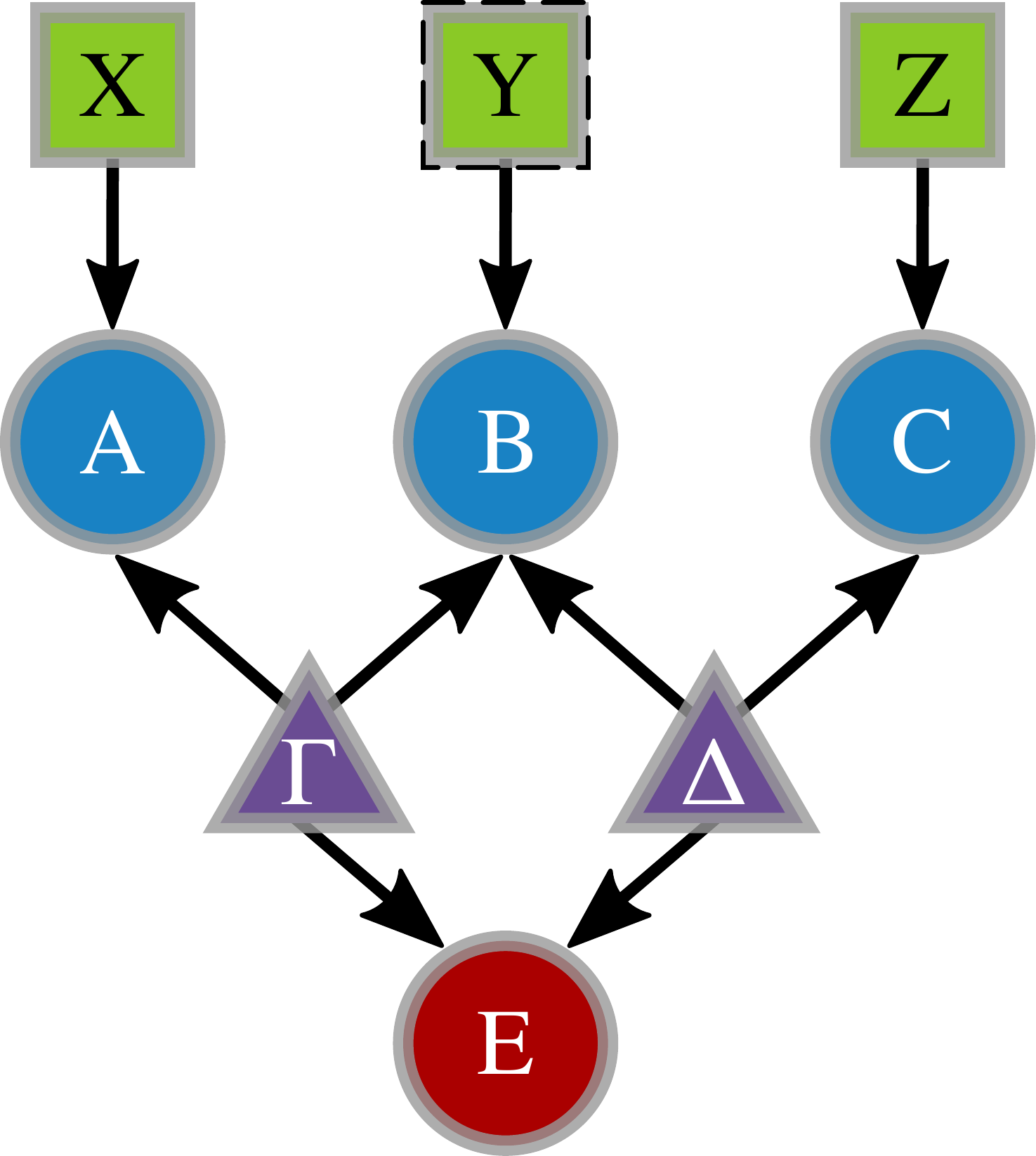}}}
\hspace{2cm}
\subfloat[]{\label{fig:bilocal_dag_ΛE}{\includegraphics[width=0.2\textwidth]{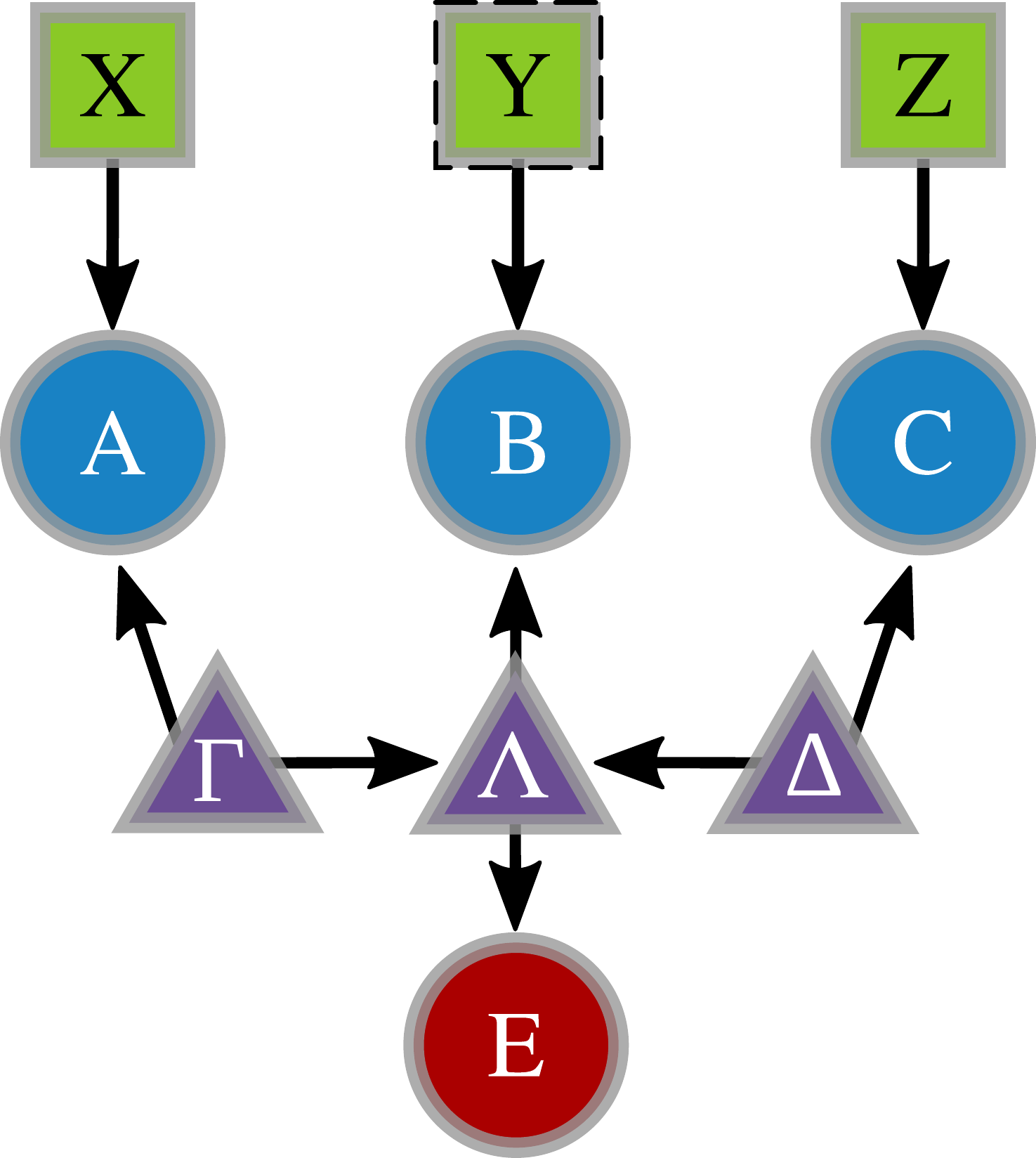}}} 
\caption{\textbf{Representation of different Eavesdropping scenarios.} We reproduce here for convenience the DAGs representing the three possible eavesdropping models in the Bilocal scenario presented in the main text: the double eavesdropper (DE), the weak eavesdropper (WE), and the strong-eavesdropper (SE) scenario.}
\label{fig:bilocal_dags}
\end{figure}

\subsection{Strong eavesdropper scenario}
In this section, we will focus on the \emph{strong-eavesdropper} (SE) scenario. 
Eve can access both sources at the same time through a central latent node as shown in Fig~\ref{fig:bilocal_dags}.

We will study different strategies, based on the protocol performed by the nodes.

\subsubsection{Bob's perform a Bell-state measurement (case 14)}

First, we provide a minimal model that explains the strategy adopted by Eve when Bob performs a Bell-state measurement. \\

The vulnerability of the protocol is evident in this scenario since Bob only performs one measurement. This leaves Eve free to project onto the eigenbasis of Bob's measurement, i.e. the Bell basis, maximizing its guessing probability of Bob's node up to unity. Crucially, this operation is also compatible with the specific causal structure under analysis. The resulting state reads
\eq{
\label{eq:rho_tilde_14_Eve}
\begin{array}{lll}
\tilde \rho  = \dfrac{1}{4}\Big(&

\ket{\Phi^+_{AC}}\bra{\Phi^+_{AC}} \;
\otimes \;
\ket{\Phi^+_{B}}\bra{\Phi^+_{B}}

\;\;\;+ \;\;\;

\ket{\Phi^-_{AC}}\bra{\Phi^-_{AC}} \;
\otimes \;
\ket{\Phi^-_{B}}\bra{\Phi^-_{B}} &\\\\

&
 + \;\;\; \ket{\Psi^+_{AC}}\bra{\Psi^+_{AC}} \;
\otimes \;
\ket{\Psi^+_{B}}\bra{\Psi^+_{B}}

\;\;\;+ \;\;\;

\ket{\Psi^-_{AC}}\bra{\Psi^-_{AC}} \;
\otimes \;
\ket{\Psi^-_{B}}\bra{\Psi^-_{B}}

&
\Big),

\end{array}
}
where we used the decomposition 

\eq{
\ket{\Psi^-_{AB_1}} 
\ket{\Psi^-_{B_2C}} = 
\dfrac{1}{2}\left(
-\ket{\Phi^+_{AC}} 
\ket{\Phi^+_{B}}
+
\ket{\Phi^-_{AC}} 
\ket{\Phi^-_{B}}
+
\ket{\Psi^+_{AC}} 
\ket{\Psi^+_{B}}
-
\ket{\Psi^-_{AC}} 
\ket{\Psi^-_{B}}
\right).
}

This new state clearly satisfies $P_Q(a, b_0,b_1,c|\rho,x,z)=P_Q(a,b_0,b_1,c|\tilde \rho,x,z) $.  
Now, Eve can guess Bob's outcome with unit probability. What about the outcomes of Alice and Charlie? For those nodes, we can use the ``informed guess" scheme conditioned on Eve's outcome. We have

\eq{
\begin{array}{lll}
P_Q(a\vec b c|\Phi_+,x=0,z=0) &=& \dfrac{1}{4}[1+(-1)^{a+c}]\delta_{b_0}^0\delta_{b_1}^0 ,\\\\
P_Q(a\vec b c|  \Phi_- ,0,0)
 &=&\dfrac{1}{4}\delta_{b_0}^0\delta_{b_1}^1,\\\\
 P_Q(a\vec b c| \Psi_+ ,0,0)  &=& \dfrac{1}{4}\delta_{b_0}^1\delta_{b_1}^0 ,\\\\
P_Q(a\vec b c|\Psi_- ,0,0) &=& \dfrac{1}{4}[1-(-1)^{a+c}]\delta_{b_0}^1\delta_{b_1}^1.
\end{array}
}

Using Eq.\eqref{eq:intro_info_guess_advanced}, we get that $\max_{a\vec b c}P_Q(a\vec b c|\psi,0,0)$ is $1/2$ when $\psi = \Phi_+, \Psi_-$ and $1/4$ otherwise. Averaging over the four Bell states, we obtain

\eq{
P_{\rm guess}^{(14)}(\mathcal A_0,\mathcal B^{(14)}, \mathcal C_0 ) \geq P_{\rm info}(\mathcal A_0,\mathcal B^{(14)}, \mathcal C_0 ) = 0.375,\\\\
P_{\rm guess}^{(14)}(\mathcal A_0,   \mathcal C_0 ) \geq P_{\rm info}(\mathcal A_0,   \mathcal C_0 ) = 0.375.
}
In terms of certifiable bits, one has:
\eq{
H_{\rm min}^{(14)}(\mathcal A_0,\mathcal B^{(14)},\mathcal C_0) \lesssim 1.41,\;\;\;\;\;H_{\rm min}^{(14)}(\mathcal A_0,   \mathcal C_0 ) \lesssim 1.41.
}

Eve's specific strategy consists of predicting Bob's outcome with certainty, then performing an informed guess on Alice's and Charlie's bits. This explains why $H_{\rm min}^{(14)}(\mathcal A_0,\mathcal B^{(14)},\mathcal C_0) = H_{\rm min}^{(14)}(\mathcal A_0,   \mathcal C_0 ) \lesssim 1.41$. Similarly, it also explains why Eve's guessing probability for the outmost nodes (Alice and Charlie) is not improved compared to the basic strategy of Eq.\eqref{eq:bilo_info_guess_basic_bounds}.

\subsubsection{Bob performs separable measurements (case 22)}
\label{sec:case_22_E_basic}
In this case, we can use the fact that the two operators performed by Bob commute 
\eq{
[\mathcal B^{(22)}_0,\mathcal B^{(22)}_1]=[\sigma_z\otimes \sigma_z,\sigma_x\otimes\sigma_x]=0,
}
to apply a strategy similar to the case of the Bell-state measurement. Specifically, due to the commutation relations the two operators share an eigenbasis, which is given by the Bell basis itself. Eve can thus apply the same strategy previously discussed, by performing a Bell-state measurement and feeding Alice, Bob, and Charlie with the state $\tilde \rho_4$ of Eq.~\eqref{eq:rho_tilde_14_Eve} rather than the initial state $\rho_4$. 
In this way, Eve will automatically know the outcome of Bob's measurement and, at the same time, will not change the results of Alice, Bob, and Charlie's measurements. Analogously to the previous case, using Eq.~\eqref{eq:intro_info_guess_advanced}, we get:
\eq{
\label{eq:bilo_Hmin22}
H_{\rm min}^{(22)}(\mathcal A_0,\mathcal B^{(22)}_0,\mathcal C_0) \lesssim 1.41,\;\;\;\;\;H_{\rm min}^{(22)}(\mathcal A_0,   \mathcal C_0 ) \lesssim 1.41.
}

\subsection{Double Eavesdropper scenario}
In this section, we comment on the causal structure shown in Fig.~\ref{fig:bilocal_dag_EF}, i.e. the ``double-eavesdropper" scenario. 
This corresponds to assuming an additional constraint on Eve, who now cannot act jointly on the qubits produced by the two sources.

The main strategy described in the previous sections consists of Eve projecting Bob's qubits on the Bell basis, which strongly relies on the ability to access both sources at the same time. When the eavesdropper can only access the sources separately, that strategy becomes unavailable, explaining why a higher number of random bits can be certified. 
Specifically, the numerical optimization confirms that in the case of maximal visibility $\nu=1$, Eve's best strategy consists of simply using the ``informed guess" protocol of Eq.\eqref{eq:intro_info_guess_basic}, thus leading to the certified bits reported in Eq.\eqref{eq:bilo_info_guess_basic_bounds}, which coincide with the results of the numerical simulations, confirming that this is indeed the optimal strategy.

\subsection{Weak Eavesdropper randomness using self-testing}
Self-testing protocols for networks, and specifically for the bilocality scenario, have been demonstrated in ref.\cite{renou2018self}
Using this idea, we can certify that a specific quantum strategy was employed in the network, based only on the observable probability distribution.
This in turn will allow us to exclude the possibility of an active eavesdropper strategy accessing the sources, leaving the \emph{informed guess} strategy as the optimal one. 
We will first consider the tilted strategy described in Eq.13 of the main text.

Consider the bilocality scenario with dichotomic measurements $A_x^a \in \mL(\Hil_A)$ and $C_z^c \in \mL(\Hil_C)$, and a four-outcome measurement $B^b_{B_1 B_2} \in \mL(\Hil_{B_1} \otimes \Hil_{B_2})$, and let us call $\expval{A_x C_y}_b = \sum_{ac} (-1)^{a+c} p(a,c|b)$ the correlator on $A$ and $C$ conditioned on having outcome $b$ for $B$.
Similarly, we can define the postselected state $\rho^b_{AC} \in \mL(\Hil_{A} \otimes \Hil_{C})$, as the effective state shared by $A$ and $C$ for each outcome $b$ of the central node.
We will start by stating the following lemma.
\begin{lemma}
\label{lm:delta_selftest}
Assume that for a given $b$ we have
    \begin{align}
        &\expval{A_0 C_0}_b = 0
        &\expval{A_1 C_0}_b = \expval{A_0 C_1}_b = \cos \delta 
        &&\expval{A_1 C_1}_b = -\sin 2\delta
        \label{eq:delta_AC_corr}
    \end{align}
and $\expval{A_0}_b = \expval{A_1}_b = \expval{C_0}_b = \expval{C_1}_b = 0$. Then, up to local unitary $U_A \otimes V_C$, the postselected state is $\rho^b_{AC} = \ket{\Phi^+_{A'C'}}\!\bra{\Phi^+_{A'C'}} \otimes \rho_{\mathrm{junk}}$ and the corresponding measurements on $A$ and $C$ are of the form $ A_x \otimes \eye_\mathrm{junk}$ and $C_z \otimes \eye_\mathrm{junk}$ respectively, where
\begin{align}
    \nonumber
    &A_0 = \sigma_z \quad A_1 = \sigma_x \cos \delta - \sigma_z \sin \delta\\
    &C_0 = \sigma_x \quad C_1 = \sigma_z \cos \delta - \sigma_x \sin \delta
    \label{eq:delta_AC_meas}
\end{align}
\end{lemma}

\begin{proof}
    A proof can be derived directly from the self-testing result contained in section D of the Supplemental Material contained in ref.\cite{wooltorton2022tight}, noticing that Eq.\eqref{eq:delta_AC_corr} maximizes the inequality $I_\delta$. 
\end{proof}

Applying slight variations of Lemma~\ref{lm:delta_selftest} repeatedly, we can obtain self-testing results for each Bell state $\{\ket{\Phi^b}\}_b = \{\ket{\Phi^+}, \ket{\Psi^+}, \ket{\Phi^-}, \ket{\Psi^-}\}$, based on having postselected correlations of the form:
\begin{align}
    \nonumber
    &\expval{A_0 C_1}_{b=(b_0, b_1)} = (-1)^{b_0} \cos \delta 
    &\expval{A_1 C_0}_{b=(b_0, b_1)} = (-1)^{b_1} \cos \delta\\
    &\expval{A_0 C_0}_b = 0
    &\expval{A_1 C_1}_{b=(b_0, b_1)} = \delta_{b_0,b_1}(-1)^{b_0}\sin(2\delta)
    \label{eq:delta_AC_corr_b}
\end{align}
By using this we can conclude the following:
\begin{lemma}
    \label{lm:bilo_selftest}
    Given a bilocality scenario with a four-outcome measurement $B^b$, measurements $A_x$ and $C_z$ satisfying Eq.\eqref{eq:delta_AC_corr_b}, there exist two completely positive and unital maps $\mC_1: \mL(\Hil_{B_1}) \to \mL(\Hil_{B'_1}), \mC_2: \mL(\Hil_{B_2}) \to \mL(\Hil_{B'_2})$, such that $(\mC_1 \otimes \mC_2) (B^b_{B_1 B_2}) = \ket{\Phi^b_{B_1' B_2'}}\!\bra{\Phi^b_{B_1' B_2'}}$.
    
    Moreover, there exist local maps $\mL_1: \mL(\Hil_{A B_1}) \to \mL(\Hil_{A' B_1})$ and $\mL_2: \mL(\Hil_{B_2 C}) \to \mL(\Hil_{B_2' C'})$ such that $\mL_1(\rho_{A B_1}) = \ket{\Phi^+_{A' B_1'}}\!\bra{\Phi^+_{A' B_1'}} \otimes \rho_\mathrm{junk}$ and $\mL_2(\sigma_{B_2 C}) = \ket{\Phi^+_{B_2' C'}}\!\bra{\Phi^+_{B_2' C'}} \otimes \rho_\mathrm{junk}$
    where $\rho_{A B_1}$ and $\sigma_{B_2 C}$ are the initial states of the two sources.
\end{lemma}
\begin{proof}
    To prove it, we can use the fact that having a distribution like the one in Eq.\eqref{eq:delta_AC_corr_b} self-tests for a strategy with measurements that are unitarily equivalent to Eq.\eqref{eq:delta_AC_meas} and states such that $U_A\otimes V_C \rho_{AC}^b U_A^\dagger \otimes V_C^\dagger = \ket{\Phi^b_{AC}}\!\bra{\Phi^b_{AC}} \otimes \rho_\mathrm{junk}$ for some local unitary $U_A\otimes V_C$.
    Following ref.~\cite{renou2018self}, we can then define the maps $\mathcal{C}_1, \mathcal{C}_2$ as the ones generated by the states $\alpha_{B_1' B_1} = \Tr_\mathrm{junk}(U_A \rho_{A B_1} U_A^\dagger)$ and $\gamma_{B_2 B_2'} =\Tr_\mathrm{junk}(U_C \rho_{B_2 C} U_C^\dagger)$.
    Indeed, when applied to $B^b_{B_1 B_2}$ we have
    \begin{multline}
        (\mathcal{C}_1 \otimes \mathcal{C}_2) (B^b_{B_1 B_2}) =
        \Tr_{B_1 B_2}(\alpha_{B'_1 B_1} \gamma_{B_2 B'_2} \; B^b_{B_1 B_2}) = \\
        \Tr_{\mathrm{junk}}\left( U_{AC} \Tr_{B_1 B_2}(\rho_{A B_1} \rho_{B_2 C} \; B^b_{B_1 B_2}) U_{AC}^\dagger\right) = 
        \ket{\Phi^b_{B_1' B_2'}}\!\bra{\Phi^b_{B_1' B_2'}}
    \end{multline}
    where, to simplify the notation we are using only the subscripts to keep track of the spaces the operators act on.
    Also, it is direct to see that $\mathcal{C}_1, \mathcal{C}_2$ are CP since $\alpha_{B_1' B_1}$ and $\gamma_{B_2 B_2'}$ are positive operators, and it is unital since $\mathcal{C}_1(\eye_{B_1}) = \Tr_{B_1}(\alpha_{B_1' B_1}) = \sum_b \Tr_{B_1 B_2 B_2'}(\alpha_{B_1' B_1} \gamma_{B_2 B_2'} B^b_{B_1 B_2}) = \sum_b \Tr_{B_2'}\left(\ket{\Phi^b_{B_1' B_2'}}\!\bra{\Phi^b_{B_1' B_2'}}\right) = \eye_{B_1'}$, and similarly for $\mathcal{C}_2$.

    To prove that the source states are equivalent to Bell states call $\mB_1$ and $\mB_2$ the maps associated with the CJ operators $\Lambda_1 = \Tr_{B_2} (V_{B_1'} B^0_{B_1 B_2}\sigma_{B_2 B_1'} V_{B_1'}^\dagger)$ and $\Lambda_2 = \Tr_{B_1} (U_{B_2'} B^0_{B_1 B_2}\rho_{B_2' B_1} U_{B_2'}^\dagger)$ respectively, and we define the maps $\mL_1 ,\mL_2 $ as $\mL_1 = \mU \otimes \mB_1$ and $\mL_2 = \mB_2 \otimes \mV$, where $\mU$ and $\mV$ are the unitary transformations given by $U_A$ and $V_C$ respectively. 
    In this way, we have that:
    \begin{multline}
        \mL_1(\rho_{A B_1}) = 
        \Tr_{B_1}\left( U_A \rho_{A B_1} U_A^\dagger {\Lambda_1}_{B_1 B_1'} \right) =
         U_A V_{B_1'} \Tr_{B_1 B_2}\left(\rho_{A B_1} B^0_{B_1 B_2}\sigma_{B_2 B_1'} \right) U_A^\dagger  V_{B_1'}^\dagger = \\
         U_A V_{B_1'} \rho^0_{A B_1'} U_A^\dagger V_{B_1'}^\dagger =
        \ket{\Phi^0_{A'B_1'}}\!\bra{\Phi^0_{A'B_1'}} \otimes \rho_\mathrm{junk}
    \end{multline}
    and similarly for $\mL_2(\sigma_{B_2 C})$. 
\end{proof}

Using the lemma above, we can conclude that the optimal strategy corresponds to the one we called \emph{informed guess}.
\begin{prop}
    For the scenarios represented by the DAGs in Fig.\ref{fig:bilocal_dag_EF} -\ref{fig:bilocal_dag_E}, if the observed distribution satisfies Eq.\eqref{eq:delta_AC_corr_b}, the optimal strategy for the eavesdropper gives $H_{\min }(ABC|E x=0, z=0) = -\log(\max p(abc|x=0, z=0))$, and $H_{\min }(AC|E x=0, z=0) = -\log(\sum_b p(b) \max p(ac|b,x=0, z=0))$.
\end{prop}
\begin{proof}
    Using lemma~\ref{lm:bilo_selftest} we have that $(\mL_1 \otimes \mL_2)(\rho_{ABC}) = \ket{\Phi^+_{A B_1}}\!\bra{\Phi^+_{A B_1}}\otimes \ket{\Phi^+_{B_2 C}}\!\bra{\Phi^+_{B_2 C}} \otimes \rho_\mathrm{junk}$, where $= \rho_{ABC} = \Tr_E(\rho_{ABCE})$.
    This means that we have $\rho_{ABCE} = \rho_{ABC} \otimes \rho_E$, and the eavesdropper cannot acquire any information by measuring her subsystem.
\end{proof}

This result indeed coincides with the maximum found with the numerical optimization in the case of the DE scenario (see Fig.6 in the main text), but it is also valid for the WE scenario, for which the optimization technique we employed cannot be used.
Specifically, this allows us to reach $4$ bits of certified randomness in both cases.